\documentclass[12pt, onecolumn, draftclsnofoot]{IEEEtran}%
\usepackage[T1]{fontenc}
\usepackage{amsmath,amssymb,amsfonts,mathrsfs,bm}% Typical maths resource packages
\usepackage{mathtools}
\usepackage{amsthm}
\usepackage{cite}
\usepackage[shortlabels]{enumitem}
\usepackage{graphicx}
\usepackage{epstopdf}
\usepackage{url}
\usepackage{colortbl}
\usepackage{multirow}
\usepackage{xcolor}
\usepackage[normalem]{ulem}
\usepackage{array}
\newcolumntype{L}[1]{>{\raggedright\let\newline\\\arraybackslash\hspace{0pt}}m{#1}}
\newcolumntype{C}[1]{>{\centering\let\newline\\\arraybackslash\hspace{0pt}}m{#1}}
\newcolumntype{R}[1]{>{\raggedleft\let\newline\\\arraybackslash\hspace{0pt}}m{#1}}

\makeatletter
\let\MYcaption\@makecaption
\makeatother
\usepackage[font=footnotesize]{subcaption}
\makeatletter
\let\@makecaption\MYcaption
\makeatother

\usepackage{xparse}



\usepackage{glossaries}
\newacronym{wrt}{w.r.t.}{with respect to}
\newacronym{RHS}{R.H.S.}{right-hand side}
\newacronym{LHS}{L.H.S.}{left-hand side}
\newacronym{iid}{i.i.d.}{independent and identically distributed}
\usepackage{float}

\ifx\notloadhyperref\undefined
	\ifx\loadbibentry\undefined
		\usepackage[hidelinks,hypertexnames=false]{hyperref} 
	\else
		\usepackage{bibentry}
		\makeatletter\let\saved@bibitem\@bibitem\makeatother
		\usepackage[hidelinks,hypertexnames=false]{hyperref}
		\makeatletter\let\@bibitem\saved@bibitem\makeatother
	\fi
\else
	\relax
\fi

\usepackage[capitalize]{cleveref}
\crefname{equation}{}{}
\Crefname{equation}{}{}
\crefname{claim}{claim}{claims}
\crefname{step}{step}{steps}
\crefname{line}{line}{lines}
\crefname{condition}{condition}{conditions}
\crefname{dmath}{}{}
\crefname{dseries}{}{}
\crefname{dgroup}{}{}
\crefname{Theorem}{Theorem}{Theorems}
\crefname{Corollary}{Corollary}{Corollaries}
\crefname{Proposition}{Proposition}{Propositions}
\crefname{Lemma}{Lemma}{Lemmas}
\crefname{Definition}{Definition}{Definitions}
\crefname{Example}{Example}{Examples}
\crefname{Assumption}{Assumption}{Assumptions}
\crefname{Remark}{Remark}{Remarks}
\crefname{Rem}{Remark}{Remarks}
\crefname{remarks}{Remarks}{Remarks}
\crefname{Exercise}{Exercise}{Exercises}
\crefname{Theorem_A}{Theorem}{Theorems}
\crefname{Corollary_A}{Corollary}{Corollaries}
\crefname{Proposition_A}{Proposition}{Propositions}
\crefname{Lemma_A}{Lemma}{Lemmas}
\crefname{Definition_A}{Definition}{Definitions}

\usepackage{algorithm,algorithmic}

\ifx\loadbreqn\undefined
	\relax
\else
	\usepackage{breqn} 
\fi

\ifx\renewtheorem\undefined
\ifx\useTheoremCounter\undefined
\newtheorem{Theorem}{Theorem}
\newtheorem{Corollary}{Corollary}
\newtheorem{Proposition}{Proposition}
\newtheorem{Lemma}{Lemma}
\else
\newtheorem{Theorem}{Theorem}
\newtheorem{Corollary}[theorem]{Corollary}
\newtheorem{Proposition}[theorem]{Proposition}
\fi

\newtheorem{Definition}{Definition}
\newtheorem{Example}{Example}

\newtheorem{Assumption}{Assumption}

\fi

\theoremstyle{remark}

\theoremstyle{plain}

\newcommand{\Real}{\mathbb{R}}

\newcommand{\iu}{\mathfrak{i}\mkern1mu}

\newcommand{\calG}{\mathcal{G}}
\newcommand{\calH}{\mathcal{H}}

\newcommand{\calV}{\mathcal{V}}

\newcommand{\bbC}{\mathbb{C}}

\newcommand{\bbZ}{\mathbb{Z}}

\DeclareSymbolFont{bsfletters}{OT1}{cmss}{bx}{n}
\DeclareSymbolFont{ssfletters}{OT1}{cmss}{m}{n}
\DeclareMathSymbol{\bsfGamma}{0}{bsfletters}{'000}
\DeclareMathSymbol{\ssfGamma}{0}{ssfletters}{'000}
\DeclareMathSymbol{\bsfDelta}{0}{bsfletters}{'001}
\DeclareMathSymbol{\ssfDelta}{0}{ssfletters}{'001}
\DeclareMathSymbol{\bsfTheta}{0}{bsfletters}{'002}
\DeclareMathSymbol{\ssfTheta}{0}{ssfletters}{'002}
\DeclareMathSymbol{\bsfLambda}{0}{bsfletters}{'003}
\DeclareMathSymbol{\ssfLambda}{0}{ssfletters}{'003}
\DeclareMathSymbol{\bsfXi}{0}{bsfletters}{'004}
\DeclareMathSymbol{\ssfXi}{0}{ssfletters}{'004}
\DeclareMathSymbol{\bsfPi}{0}{bsfletters}{'005}
\DeclareMathSymbol{\ssfPi}{0}{ssfletters}{'005}
\DeclareMathSymbol{\bsfSigma}{0}{bsfletters}{'006}
\DeclareMathSymbol{\ssfSigma}{0}{ssfletters}{'006}
\DeclareMathSymbol{\bsfUpsilon}{0}{bsfletters}{'007}
\DeclareMathSymbol{\ssfUpsilon}{0}{ssfletters}{'007}
\DeclareMathSymbol{\bsfPhi}{0}{bsfletters}{'010}
\DeclareMathSymbol{\ssfPhi}{0}{ssfletters}{'010}
\DeclareMathSymbol{\bsfPsi}{0}{bsfletters}{'011}
\DeclareMathSymbol{\ssfPsi}{0}{ssfletters}{'011}
\DeclareMathSymbol{\bsfOmega}{0}{bsfletters}{'012}
\DeclareMathSymbol{\ssfOmega}{0}{ssfletters}{'012}

\DeclareMathOperator*{\argmin}{arg\,min}

\DeclareMathOperator{\spn}{span}

\DeclareMathOperator{\sinc}{sinc}

\newcommand{\qednew}{\nobreak \ifvmode \relax \else
      \ifdim\lastskip<1.5em \hskip-\lastskip
      \hskip1.5em plus0em minus0.5em \fi \nobreak
      \vrule height0.75em width0.5em depth0.25em\fi}

\newcommand{\ud}{\mathrm{d}}
\newcommand{\Id}{\mathrm{Id}}

\newcommand{\ofrac}[1]{{\frac{1}{#1}}}

\newcommand{\tc}[1]{^{(#1)}}

\newcommand{\floor}[1]{\left\lfloor{#1}\right\rfloor}
\newcommand{\ip}[2]{{\left\langle{#1},\, {#2}\right\rangle}}
\newcommand{\norm}[1]{{\left\lVert{#1}\right\rVert}}

\newcommand{\cond}[2]{\left. {#1}\, \middle| \, {#2} \right.}

\DeclareDocumentCommand \P { g d() g } {%
	\IfNoValueTF {#3} 
	{%
		\IfNoValueTF {#1} 
		{%
			\IfNoValueTF {#2}
			{%
				\mathbb{P}%
			}%
			{%
				\mathbb{P}\left({#2}\right)%
			}%
		}%
		{%
			\IfNoValueTF {#2}
			{%
				\mathbb{P}_{#1}%
			}%
			{%
				\mathbb{P}_{#1}\left({#2}\right)%
			}%		
		}%
	}%
	{%
		\IfNoValueTF {#1} 
		{%
			\mathbb{P}\left(\cond{#2}{#3}\right)%
		}%
		{%
			\mathbb{P}_{#1}\left(\cond{#2}{#3}\right)%
		}%	
	}%
}

\DeclareDocumentCommand \E { g o g } {%
	\IfNoValueTF {#3} 
	{%
		\IfNoValueTF {#1} 
		{%
			\IfNoValueTF {#2}
			{%
				\mathbb{E}%
			}%
			{%
				\mathbb{E}\left[{#2}\right]%
			}%
		}%
		{%
			\IfNoValueTF {#2}
			{%
				\mathbb{E}_{#1}%
			}%
			{%
				\mathbb{E}_{#1}\left[{#2}\right]%
			}%		
		}%
	}%
	{%
		\IfNoValueTF {#1} 
		{%
			\mathbb{E}\left[\cond{#2}{#3}\right]%
		}%
		{%
			\mathbb{E}_{#1}\left[\cond{#2}{#3}\right]%
		}%	
	}%
}

\definecolor{gray90}{gray}{0.9}

\ifx\nohighlights\undefined

	\newcommand{\msout}[1]{\text{\color{green} \sout{\ensuremath{#1}}}}
	\newcommand{\del}[1]{{\color{green}\ifmmode \msout{#1}\else\sout{#1}\fi}}
\else

	\newcommand{\msout}[1]{#1}
	\newcommand{\del}[1]{#1}
\fi

\newcommand{\hide}[1]{}
\renewcommand{\figurename}{Fig.}
\newcommand{\figref}[1]{\figurename~\ref{#1}}
\graphicspath{{./Figures/}} 
\ifx\diagnoselabel\undefined
	\relax
\else
	\makeatletter
	 \def\@testdef #1#2#3{%
		 \def\reserved@a{#3}\expandafter \ifx \csname #1@#2\endcsname
		\reserved@a  \else
	 \typeout{^^Jlabel #2 changed:^^J%
	 \meaning\reserved@a^^J%
	 \expandafter\meaning\csname #1@#2\endcsname^^J}%
	 \@tempswatrue \fi}
	\makeatother
\fi

\usepackage{tikz}
\usepackage{color}
\usepackage{pgfplots}
\usepackage{graphicx}
\pgfplotsset{compat=1.5}%%%%%%%%%%%%%%%%%%%

\graphicspath{{./Figures/}} 
\providecommand{\U}[1]{\protect\rule{.1in}{.1in}}
\theoremstyle{definition}

\title{A Hilbert Space Theory of Generalized Graph Signal Processing}

\author{
Feng~Ji and Wee~Peng~Tay,~\IEEEmembership{Senior Member,~IEEE}%
\thanks{This research is supported by the Singapore Ministry of Education Academic Research Fund Tier 2 grant MOE2018-T2-2-019.}% 
\thanks{The authors are with the School of Electrical and Electronic Engineering, Nanyang Technological University, Singapore (e-mail: jifeng@ntu.edu.sg, wptay@ntu.edu.sg). }
}

\begin{document}

\maketitle
\begin{abstract}
Graph signal processing (GSP) has become an important tool in many areas such as image processing, networking learning and analysis of social network data. In this paper, we propose a broader framework that not only encompasses traditional GSP as a special case, but also includes a hybrid framework of graph and classical signal processing over a continuous domain. Our framework relies extensively on concepts and tools from functional analysis to generalize traditional GSP to graph signals in a separable Hilbert space with infinite dimensions. We develop a concept analogous to Fourier transform for generalized GSP and the theory of filtering and sampling such signals.  
\end{abstract}


\section{Introduction}

Since its emergence, the theory and applications of graph signal processing (GSP) have rapidly developed (see for example, \cite{Shu13, San13, San14, Gad14, Don16, Egi17, Sha17, Gra18, Ort18, Girault2018}). Traditional GSP theory is essentially based on a change of orthonormal basis in a finite dimensional vector space. Suppose $G=(V,E)$ is a weighted, undirected graph with $V$ the vertex set of size $n$ and $E$ the set of edges. Recall that a graph signal $f$ assigns a complex number to each vertex, and hence $f$ can be regarded as an element of $\bbC^n$, where $\bbC$ is the set of complex numbers. The heart of the theory is a shift operator $A_G$ that is usually defined using a property of the graph. Examples of $A_G$ include the adjacency matrix and the Laplacian matrix of $G$. Graph shift operators are typically chosen to be symmetric. By the spectral theorem, all the eigenvalues of $A_G$ are real and $\bbC^n$ has an orthonormal basis consisting of eigenvectors of $A_G$. Therefore, one can define the notion of vertex and frequency domains, on which the rest of the theory builds upon.

%\begin{figure}[!t] 
%\centering
%\includegraphics[width=0.7\columnwidth]{1}
%\caption{Generalized graph signals: each vertex may be associated with an $L^2$ function over a finite closed interval.} \label{fig:1}
%\end{figure}

However, instead of assigning a complex number at each vertex, one can assign mathematical objects with richer structures to each vertex of a graph. An example of such a mathematical object comes from a Hilbert space. In particular, it would be of interest to assign an $L^2$ function over a finite closed interval $[a,b]$ to each vertex of a graph. Such a consideration is not just a plain generalization, as it has important practical applications in for example, sensor networks and social networks, where each node in the network is observing a time-varying continuous signal. We have the following considerations.
\begin{Example} \label{eg:main}\ 
\begin{enumerate}[(a)]
	\item The case where each vertex signal is a complex number belongs to the traditional GSP framework, which has been extensively studied in \cite{Shu13, San13, San14, Gad14, Don16, Egi17, Sha17, Gra18, Ort18, Girault2018}.
	\item An extension of traditional GSP to the case where each vertex signal is a finite-length \emph{discrete} time series is proposed in the time-vertex GSP framework \cite{Gra18}. Such a vertex signal is from a Hilbert space $\bbC^m$ for some $m\geq 1$. The assumption here is that all the time series at different vertices share the same time indices. 
	\item \label{it:1} The case where the signal at each graph vertex $v\in V$ is itself a graph signal on a finite graph $G_v$ has not been studied in the literature, to the best of our knowledge. In this case, each vertex signal is from $\bbC^m$ for some $m\geq1$, but unlike the time-vertex GSP framework, the underlying graph topology can be more general (the time-vertex GSP framework is equivalent to using a path graph with $m$ vertices). Furthermore, different vertices can have different underlying graphs for their signals, and the correspondence between the vertices in $G_v$ and those in $G_u$ for $v\ne u$ need not be one-to-one. 
	
%<*tag:air>
	Applications include relating the signals in one social network to another social network, or in problems where the underlying graph at each vertex of $G$ is varying (cf.\ Section~\ref{subsec:adaptive}). For example, when studying signals on time-varying adaptive networks (\cite{Ito02, Bar08, Zsc12, Gro09}), which includes social networks, biological networks and neural networks, we may let $G$ be a path graph representing a finite portion of the time line. The signal $f_t$ at each node $t$ is a vector in $\mathbb{C}^m$. Due to the additional local structures in the signal, it is usually advantageous to consider these structures when performing signal analysis. Thus, we can regard $f_t$ as a graph signal corresponding to time $t$. Furthermore, multiple graph signals $f_t$ from different networks can be related to each other, forming yet another underlying graph at each time~$t$.
	
	\item \label{it:2} The case where the signal at each vertex is drawn from an infinite dimensional Hilbert space has again never been studied. An example is continuous time signals that are not bandlimited (in the time direction). In this case, from the Shannon-Nyquist Theorem \cite{Sha49}, it is impossible to recover the full undistorted signal using a finite sampling rate (cf.\ Section~\ref{subsec:inf_prop}). Thus, the time-vertex GSP framework, which requires discrete time series at all graph vertice, would introduce errors in the inference procedure. Furthermore, signals may not be sampled synchronously in a sensor network (sensor synchronization usually requires additional effort \cite{Siv04, Ols10, Bru17}) so that the time-vertex GSP framework may not be the best approach. A more general GSP framework is thus needed to address many practical engineering scenarios.
			
In addition, a Hilbert space, such as the space of $L^2$ functions over an appropriate domain $\Omega$, usually has rich internal structures. The usual GSP framework takes a ``snapshot picture" by looking at the graph signal at $x \in \Omega$ one by one. This approach may easily disregard the internal relations among the points in $\Omega$. In this paper, the proposed framework avoids such a local consideration entirely, and fuses the graph operators with operations on $L^2(\Omega)$ for signal processing to achieve minimal information loss. For example, to study continuous time graph signals mentioned above, it can be more beneficial to consider the graph signal to be a function belonging to $L^2(\mathbb{R})$, such that we may apply operations such as continuous Fourier transform and wavelet transforms in conjunction with graph based transforms. We shall demonstrate this by the example of information propagation on social networks in Section~\ref{subsec:inf_prop}.
%</tag:air>
\end{enumerate}
\end{Example}

In this paper, we propose and develop a Hilbert space theory of generalized GSP that can handle all the above cases. To do that, it is inevitable that the theory is based on an abstract foundation, which  requires the reader to have a good grasp of functional analysis concepts and tools. For these, we refer the reader to \cite{Lax02, DebMik:B05}. We shall constantly refer back to Example~\ref{eg:main} to illustrate and explain theoretical results.

We shall develop the theory parallel to classical signal processing and GSP. Important notions such as convolution filters and bandlimitedness are best understood when signals are viewed in the frequency domain. Therefore, defining what constitutes the frequency domain is a hallmark of both classical signal processing and traditional GSP. In the same spirit, we first define the frequency domain for the generalized GSP framework using the spectral theory of compact operators on Hilbert spaces. We then proceed to discuss filtering and sampling. A preliminary version of this paper was presented in the conference paper \cite{JiTay18}, which introduced some of the basic concepts in this paper without proof.

The rest of the paper is organized as follows. In Section~\ref{sec:gen}, we set up the framework by defining graph signals in a separable Hilbert space. In Section~\ref{sec:tra}, we introduce the notion of frequency. Based on this, we develop the concepts of filtering and sampling in Section~\ref{sec:fil} and Section~\ref{sec:sam}. We present numerical results in Section~\ref{sec:sim} and conclude in Section~\ref{sec:con}. Throughout, we provide examples to highlight situations where the framework of the paper is not only applicable but also necessary. 

\emph{Notations}. We use $\Real$ and $\mathbb{C}$ to denote the real and complex fields, respectively, while $\bbZ$ denotes the set of integers. The symbol $\otimes$ denotes tensor product, $\circ$ is function composition, and $\cong$ denotes isomorphic equivalence. $\Id$ is the identity operator and $\iu=\sqrt{-1}$. $\bbC^n$ is equipped with an inner product $\ip{\cdot}{\cdot}_{\bbC^n}$ with corresponding norm $\norm{\cdot}_{\bbC^n}$. The space $L^2(\Omega)=L^2(\Omega,\mathcal{F},\mu)$ with $(\Omega,\mathcal{F},\mu)$ a measure space, is the collection of functions $f:\Omega \mapsto \bbC$ such that $\int_\Omega |f|^2 \ud\mu < \infty$.

\section{Generalized graph signals} \label{sec:gen}

Let $G=(V,E)$ be a simple finite undirected weighted graph and $X$ be a metric space.  In this paper, when we talk about a vector space, we always assume that the base field is the complex numbers $\mathbb{C}$, unless otherwise stated. In traditional GSP, the signals on the vertices of $G$ are assumed to be real or complex. We generalize this as follows.

\begin{Definition}
Suppose $\mathcal{H}$ is a \emph{Hilbert space} (i.e., a complete inner product space). A \emph{graph signal in $\mathcal{H}$} is a function $f: V \to \mathcal{H}.$ The space of graph signals in $\mathcal{H}$ is denoted by $S(G,\mathcal{H}).$
\end{Definition}

With a few examples below, we demonstrate why this generalized notion is versatile. In short, Hilbert spaces include both the more familiar finite dimensional cases and also infinite dimensional cases that can represent more realistic signals in practice.

\begin{Example}\ \label{eg:sts}
\begin{enumerate}[(a)]
\item\label{it:GSP} Suppose that $\mathcal{H}=\bbC$, then $f\in S(G,\mathcal{H})$ is a traditional graph signal where $f(v)\in\bbC$ for each $v\in V$. This case has been extensively studied in \cite{Shu13} and the references therein.

\item\label{it:L2} 
%<*tag:slh>
Suppose $\mathcal{H}=L^2([a,b])$, the space of complex-valued $L^2$ functions over a finite closed interval $[a,b]$, with the Lebesgue measure (cf. Example~\ref{eg:main}\ref{it:2}). A graph signal $f$ in $\mathcal{H}$ assigns an $L^2$ function on $[a,b]$ to each vertex of $G$. Alternatively, we can view $f$ as a function assigning a complex number to each point in $V\times [a,b]$ (where $\times$ denotes the usual Cartesian product). Consequently, as $G$ is a finite graph, $S(G,\mathcal{H}) = L^2(V\times [a,b])$. If $b>a\geq 0$, $S(G,\mathcal{H})$ can be viewed as a family of traditional graph signals parametrized by a continuous time domain $[a,b]$. 
To deal with such an $\mathcal{H}$, one may use finite dimensional subspaces, e.g., polynomials of bounded degree, for approximations. However, there are continuous functions that render such a strategy a failure (cf.\ Theorems~5.4 and 5.5 in \cite{Car09}). Therefore, it is useful to have a framework that encompasses such signals from infinite dimensional Hilbert spaces.
%</tag:slh>
\item\label{it:swd} Suppose $\mathcal{H} =L^2(G')$ with discrete measure, where $G'=(V',E')$ is a finite graph (possibly directed). By finiteness, $\mathcal{H}$ is the space of finite complex graph signals on $G'$. We define the graph structure for the product $G\times G'$ as follows (see \figref{fig:Gproduct} for an example): $(v_1,u_1)$ is connected to $(v_2,u_2)$ in $V\times V'$ if either $v_1=v_2$ and $(u_1,u_2) \in E'$ with the edge weight the same as the weight of $(u_1,u_2)$, or $u_1=u_2$ and $(v_1,v_2)\in E$ with the edge weight the same as the weight of $(v_1,v_2)$. Therefore, $S(G,\mathcal{H})$ can be identified with complex signals on the graph $G\times G'$. In particular, if $G'$ is a path graph, $S(G,\mathcal{H})$ corresponds to the time-vertex GSP framework of \cite{Gra18}. Strictly speaking, as we may view $S(G,\mathcal{H})$ as traditional GSP signals on the graph $G\times G'$, this example remains within the framework of \cite{Shu13} in this respect.
\end{enumerate}

\begin{figure}[!htb] 
\centering
\includegraphics[width=0.65\columnwidth]{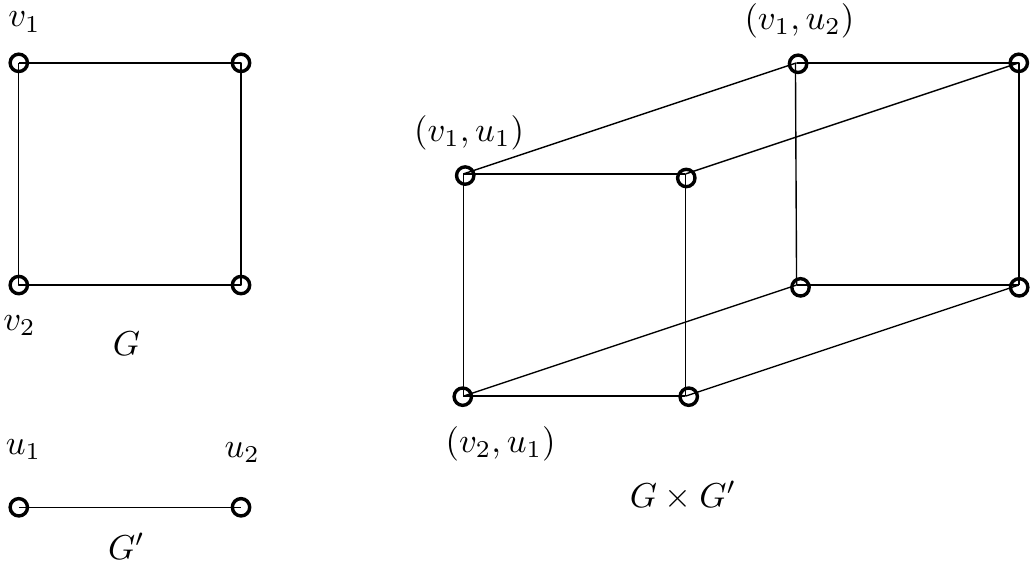}
\caption{An example of $G\times G'$.} \label{fig:Gproduct}
\end{figure}	

\end{Example}

In the following, we first review some basic facts about Hilbert spaces (cf.\ Chapter 6 of \cite{Lax02}). By definition, a Hilbert space $\mathcal{H}$ is equipped with an inner product, denoted by $\ip{\cdot}{\cdot}_\mathcal{H}$. %
%A \emph{basis} of $\mathcal{H}$ is a linearly independent collection of vectors $\Xi=\{x_{\lambda} : \lambda \in \Lambda\}$ (here $\Lambda$ is an indexing set), whose closed linear span is $\mathcal{H}$. The basis $B$ is called an \emph{orthonormal basis} if $\ip{x_{\lambda}}{x_{\lambda'}}_\mathcal{H}=\delta_{\lambda \lambda'}$, the Kronecker delta function. 
%For a Hilbert space, it is known that there is always an orthonormal basis, and any two orthonormal basis have the same cardinality. Therefore, it is meaningful to say that 
A Hilbert space is \emph{separable} if it has a countable orthonormal basis. Throughout this paper, we assume that $\mathcal{H}$ is separable, which is the case for most physical signals. Since $\mathcal{H}$ is separable, without loss of generality, we can assume $\mathcal{H} = L^2(\Omega)$ for some measurable space $\Omega$ with a measure~$\mu$ \cite[Theorem~3.4.27]{DebMik:B05}. For each $f \in S(G,\mathcal{H})$ and $u\in V$, we then have $f(u) \in L^2(\Omega)$ and $f(u)(x)\in \mathbb{C}$ for each $x\in\Omega$. A useful point of view is that we can treat $f$ as a function of two variables and write $f(u,x)$ for $f(u)(x).$ 

Recall that for a finite dimensional Euclidean vector space $\mathbb{C}^n$, we may form the \emph{tensor product} (cf.\ \cite[Chapter IV.5]{Hun03}) $\mathbb{C}^n \otimes \mathcal{H}$ as the set of finite linear sums $\sum_{i=1}^n v_i\otimes h_i,$ with $v_i\in \mathbb{C}^n$ and $h_i\in \mathcal{H}$ for all $i=1,\ldots,n$, such that the following holds for any $v_1, v_2, v\in \bbC^n$ and any $h_1, h_2, h\in \mathcal{H}$:
\begin{enumerate}[(a)]
\item $v_1\otimes h + v_2\otimes h = (v_1+v_2)\otimes h;$
\item $v\otimes h_1 + v\otimes h_2 = v\otimes (h_1+h_2);$
\item $rv\otimes h = v\otimes rh$ for $r \in \mathbb{C}$.	
\end{enumerate}
The tensor product $\mathbb{C}^n \otimes \mathcal{H}$ is equipped with an inner product $\ip{\cdot}{\cdot}_{\bbC^n\otimes\mathcal{H}}$ induced (linearly) by: 
\begin{align}\label{tensor_ip}
\ip{v_1\otimes h_1}{v_2\otimes h_2}_{\bbC^n\otimes\mathcal{H}} = \ip{v_1}{v_2}_{\mathbb{C}^n}\ip{h_1}{h_2}_{\mathcal{H}},
\end{align}
which defines a metric on $\mathbb{C}^n \otimes \mathcal{H}.$ As $\mathbb{C}^n$ is finite dimensional, $\mathbb{C}^n \otimes \mathcal{H}$ is complete and hence a Hilbert space. 
%Using the relations above, for any fixed basis $e_1,\ldots, e_n$ of $\mathbb{C}^n$, any element of $\mathbb{C}^n \otimes \mathcal{H}$ can be written as $\sum_{i=1}^n e_i \otimes h_i, h_i\in \mathcal{H}$.  
Suppose $\Phi = \{\phi_{i}\}_{1\leq i\leq n}$ is an orthonormal basis of $\mathbb{C}^n$ and $\Xi = \{\xi_j\}_{j\geq 1}$ is an orthonormal basis of $\mathcal{H}.$ Then $\Phi\otimes \Xi=\{\phi_i\otimes \xi_j\}_{1\leq i\leq n,j\geq 1}$ forms an orthonormal basis of $\mathbb{C}^n \otimes \mathcal{H}$. 

Using the tensor product construction, we have the following alternative description of $S(G,\mathcal{H})$, which shows that $S(G,\mathcal{H})$ is a separable Hilbert space.

\begin{Lemma}\label{lem:Sisomorphism}
$S(G,\mathcal{H})$ is a vector space and there is an isomorphism between $S(G,\mathcal{H})$ and $\mathbb{C}^n\otimes \mathcal{H}$ given by 
\begin{align}\label{psi}
\psi(f) = \sum_{v\in V} v \otimes f(v)
\end{align}
for $f\in S(G,\mathcal{H})$, where $V$ is identified with the standard basis of $\bbC^n$.
\end{Lemma}
\begin{IEEEproof}
The vector space structure of $S(G,\mathcal{H})$ comes from that of $\mathcal{H}$: for $f,g\in S(G,\mathcal{H})$, $u\in V$ and $r\in \mathbb{C}$, we have $(f+g)(u)=f(u)+g(u) \in \mathcal{H}$ and $(rf)(u)=rf(u)$.

It is clear that $\psi$ is an injective mapping from $S(G,\mathcal{H})$ to $\bbC^n\otimes\mathcal{H}$. For the inverse, each $a \in \bbC^n \otimes \mathcal{H}$ can be written in the form $a = \sum_{v\in V}v\otimes h_v$, where $h_v \in \mathcal{H}$. We define $\psi^{-1}(a)(v) = h_v$ so that $\psi^{-1}(a)\in S(G,H)$ and is the inverse map of $\psi$. Hence $\psi$ is an isomorphism and the proof is complete.
\end{IEEEproof}

In the rest of this paper, we will always identify $V$ with the standard basis of $\bbC^n$. Then, Lemma~\ref{lem:Sisomorphism} gives a map $\psi$ that identifies $S(G,\mathcal{H})$ with $\bbC^n\otimes \mathcal{H}$, so that we may carry structures on $\bbC^n\otimes \mathcal{H}$ to $S(G,\mathcal{H})$. In particular, $S(G,\mathcal{H})$ is a Hilbert space and we can define an inner product for $f,g\in S(G,\mathcal{H})$ as 
\begin{align}\label{S_ip}
\ip{f}{g} = \ip{\psi(f)}{\psi(g)}_{\bbC^n\otimes \mathcal{H}}.
\end{align}
Since $S(G,\mathcal{H})\cong\bbC^n\otimes\mathcal{H}$, we will often abuse notations by treating $f\in S(G,\mathcal{H})$ as an element of $\bbC^n\otimes\mathcal{H}$, keeping in mind the isomorphism map $\psi$. Furthermore, one can extend $f: V\mapsto \mathcal{H}$ uniquely to a linear transformation $\bbC^n\mapsto\mathcal{H}$, i.e., $f(u,x) = \sum_{v\in V} u(v) f(v,x)$ is uniquely defined for all $u\in\bbC^n$ and $x\in\Omega$. Here, $u(v) = \ip{u}{v}_{\bbC^n}$ is the $v$-th component of $u$ in $\bbC^n$. 

\section{\texorpdfstring{$\mathcal{F}$}{F}-transform} \label{sec:tra}

%<*tag:tga>
To give an overview, in the same spirit as traditional GSP, we introduce the notion of Fourier transformation for the generalized GSP framework. We refer to this as simply the $\mathcal{F}$-transform. In doing so, we are able to introduce the notion of a frequency domain for generalized GSP. The interplay between the graph domain and the frequency domain plays an essential role in signal processing. As signals from a Hilbert space $\mathcal{H}$ have their own transform space, we can decompose the $\mathcal{F}$-transform into simpler pieces, called partial $\mathcal{F}$-transforms, which we also introduce in this section.
%</tag:tga>

Fix orthonormal bases $\Phi$ of $\mathbb{C}^n$ and $\Xi$ of $\mathcal{H}$. We have seen that $\Phi\otimes \Xi$ forms an orthonormal basis of $S(G,\mathcal{H})$. For each $f \in S(G,\mathcal{H})$, $\phi \in \Phi$ and $\xi \in \Xi$, the joint \emph{$\mathcal{F}$-transform}  is defined as:
\begin{align}\label{F-transform}
\mathcal{F}_f(\phi\otimes \xi) = \ip{f}{\phi\otimes \xi}\triangleq\ip{\psi(f)}{\phi\otimes \xi}_{\bbC^n\otimes\mathcal{H}},
\end{align}
where $\psi$ is the isomorphism map in Lemma~\ref{lem:Sisomorphism} and the inner product on the right-hand side of \eqref{F-transform} is as defined in \eqref{tensor_ip}. Note that $\sum_{\phi\otimes\xi} |\mathcal{F}_f(\phi\otimes\xi)|^2 < \infty$ for all $f\in S(G,\mathcal{H})$.

Since we have identified $V$ with the standard basis of $\bbC^n$, we write $f(\cdot,x) = \sum_{v\in V} f(v,x) v \in \bbC^n$. The \emph{partial $\mathcal{F}$-transforms} are defined as: 
\begin{align}
\text{$\mathcal{H}$-transform:}&\ \ \mathcal{H}_f(\xi)(v) = \ip{f(v,\cdot)}{\xi}_{\mathcal{H}},\label{H-transform}\\
\text{$\calG$-transform:}&\ \ \mathcal{G}_f(\phi)(x) = \ip{f(\cdot,x)}{\phi}_{\mathbb{C}^n},\label{G-transform}
\end{align}
for every $v\in V$ and $x\in \Omega$. Since $\Phi\otimes\Xi$ is an orthonormal basis, given a sequence of numbers $g = (g(\phi\otimes \xi))_{\phi,\xi}$ such that $\sum_{\phi, \xi}|g(\phi\otimes\xi)|^2<\infty$, the \emph{inverse} $\mathcal{F}$-transform is given by 
\begin{align}
\mathcal{F}^{-1}_g = \sum_{\phi,\xi} g(\phi\otimes \xi)\cdot \phi\otimes\xi.
\end{align}
Clearly, if $g = \mathcal{F}_f$, then $\mathcal{F}^{-1}_g = f,$ and vice versa.

The definitions above do not involve the graph $G$. It appears in the following way: the orthonormal basis $\Phi$ is usually chosen as a set of eigenbasis of a symmetric graph shift operator $A_G$. Common choices of $A_G$ include the adjacency matrix and the Laplacian matrix of $G$. In the same spirit as \cite{Girault2018}, the inner product $\ip{\cdot}{\cdot}_{\bbC^n}$ and the basis $\Phi$ (hence $A_G$) can be chosen judiciously depending on the application.

\begin{Example}\ \label{eg:F-transforms}
\begin{enumerate}[(a)]
\item In Example~\ref{eg:sts}\ref{it:GSP}, $\mathcal{H}=\bbC$, which in our terminology, can be identified with $L^2(\{0\})$. In this example, we simply use $r \in \bbC$ to represent each element in $\mathcal{H}$ so that $\Xi = \{1\}$ and $\mathcal{H}_f(1)(v)=f(v)$ for each $v\in V$. The $\mathcal{H}$-transform is thus simply the graph signal itself. We also have $\calG_f(\phi) = \ip{f}{\phi}_{\bbC^n}$ for each $\phi$ in the eigenbasis of a graph shift operator. The $\calG$-transform, and hence the $\mathcal{F}$-transform, are both equivalent to the traditional graph Fourier transform in \cite{Shu13}. 

\item In Example~\ref{eg:sts}\ref{it:L2}, $\mathcal{H}=L^2([0,2\pi])$ (cf. Example~\ref{eg:main}\ref{it:2}). Using the basis $\{\exp(\iu mx)/\sqrt{2\pi}: m \in \bbZ\}$ for $\mathcal{H}$, we see that the $\mathcal{H}$-transform of $f(v)$ for each $v\in V$ is simply its Fourier transform. For each $x\in[0,2\pi]$, the $\calG$-transform of $f(\cdot,x)$ is the traditional graph Fourier transform of the graph signal $\{f(v,x) : v \in V\}$. 
%<*tag:tft>
The $\mathcal{F}$-transform however does not have any equivalence in traditional GSP (which deals with finite-dimensional spaces) as $[0,2\pi]$ is a continuous domain with $\mathcal{H}$ being infinite dimensional. For such an $\mathcal{H}$, there is an abundant family of signals (e.g., rectangular pulse functions \cite{Wan12}) whose $\mathcal{F}$-transforms cannot be described by traditional GSP Fourier transforms at finite samples.
%</tag:tft>

\item In Example~\ref{eg:sts}\ref{it:swd}, $G'=(V',E')$ is another graph and we have taken $\mathcal{H}=L^2(G')$. In this case, $S(G,\mathcal{H}) = L^2(G\times G').$ Then each $\phi\otimes \xi$ is an eigenvalue of the matrix $A_G\otimes A_{G'}$. The joint $\mathcal{F}$-transform is nothing but the graph Fourier transform of signals on $G\times G'$, which is the same as the joint Fourier transform defined in \cite{Gra18}.
\end{enumerate}
\end{Example}

Since we identify $V$ with the standard basis of $\bbC^n$, we can write $\mathcal{H}_f(\xi) = \sum_{v\in V} \mathcal{H}_f(\xi)(v) v \in \bbC^n$ for each $\xi\in\Xi$. Note also that for each $\phi\in\Phi$, $\calG_f(\phi)$ is a mapping $\Omega\mapsto\bbC$ and from the Cauchy-Schwarz inequality, we have
\begin{align*}
&\int_\Omega |\calG_f(\phi)(x)|^2 \ud\mu(x)
\leq \int_\Omega \norm{f(\cdot,x)}_{\bbC^n}^2 \ud\mu(x) \\
&= \int_\Omega \sum_{v\in V} |f(v,x)|^2 \ud\mu(x) 
= \sum_{v\in V} \int_\Omega |f(v,x)|^2 \ud\mu(x) < \infty,
\end{align*}
where the first equality follows from Parseval's formula, and the last inequality is because $f(v,\cdot)\in\mathcal{H}=L^2(\Omega)$ and $V$ is finite. Therefore, $\calG_f(\phi)\in\mathcal{H}$.

\begin{Lemma} \label{lem:fap}
For any $\phi \in \Phi$, $\xi \in \Xi$ and $f\in S(G,\mathcal{H})$, we have
\begin{align}
\mathcal{H}_f(\xi)&=\sum_{\phi'\in\Phi}\mathcal{F}_f(\phi'\otimes\xi)\phi',\label{HF}\\
\calG_f(\phi)&=\sum_{\xi'\in\Xi}\mathcal{F}_f(\phi\otimes\xi')\xi',\label{GF}
\end{align}
and
\begin{align}\label{eq:FHG}
\mathcal{F}_f(\phi\otimes \xi) = \ip{\mathcal{H}_f(\xi)}{\phi}_{\mathbb{C}^n} = \ip{\mathcal{G}_f(\phi)}{\xi}_{\mathcal{H}}.
\end{align}
\end{Lemma}

\begin{IEEEproof}
For any $\phi\in\Phi$, we have
\begin{align*}
\ip{\mathcal{H}_f(\xi)}{\phi}_{\bbC^n}
&= \ip{\sum_v \mathcal{H}_f(\xi)(v)v}{\phi}_{\bbC^n}\\
&= \sum_v \mathcal{H}_f(\xi)(v) \ip{v}{\phi}_{\bbC^n}\\
&= \sum_v \ip{f(v)}{\xi}_\mathcal{H} \ip{v}{\phi}_{\bbC^n}\ \text{ from \eqref{H-transform}}\\
&= \sum_v \ip{v\otimes f(v)}{\phi\otimes\xi}_{\bbC^n\otimes\mathcal{H}}\ \text{ from \eqref{tensor_ip}}\\
&=\ip{f}{\phi\otimes\xi}\ \text{ from \eqref{psi} and \eqref{F-transform}}\\
&=\mathcal{F}_f(\phi\otimes\xi),
\end{align*}
which proves the first equality in \eqref{eq:FHG}. Since $\Phi$ is an orthonormal basis for $\bbC^n$, \eqref{HF} follows. The other identities follow similarly and the proof is complete.
\end{IEEEproof}

As $\Phi\otimes \Xi$ is an orthonormal basis, each $f\in S(G,\mathcal{H})$ can be written as $f=\sum_{\phi\otimes \xi} \mathcal{F}_f(\phi\otimes\xi) \cdot \phi\otimes \xi$. The terms $\mathcal{F}_f(\phi\otimes\xi)$ are square summable. On the other hand, suppose we are given a mapping $L$ on $S(G,\mathcal{H})$ into $\mathbb{C}$ such that $L(\phi\otimes \xi)$ is square summable, from the inverse $\mathcal{F}$-transform, we can view $L$ as an element of $S(G,\mathcal{H})$ given by $\sum_{\phi,\xi}L(\phi\otimes \xi)\cdot \phi\otimes \xi$. An immediate example is $\mathcal{F}_f$ can be identified with $f$ when viewed as an element of $S(G,\mathcal{H})$. This point of view is useful when we discuss convolution later on.

For a general $\mathcal{H}$, an important source of orthonormal basis comes from the eigenvectors of a family of bounded linear operators. More specifically, recall that a bounded linear operator $A$ on a Hilbert space $\mathcal{H}$ is called \emph{compact} \cite[Chapter 21.1]{Lax02} if the image of the closed unit ball has a compact closure. It is moreover \emph{self-adjoint} if $\langle Ax,y \rangle_{\mathcal{H}}=\langle x,Ay \rangle_{\mathcal{H}}$ for any $x,y \in \mathcal{H}$. The spectral theorem (see \cite[Chapter 28]{Lax02}) in this case says that if $A$ is compact and self-adjoint, then all the eigenvalues of $A$ are real and $\mathcal{H}$ has an orthonormal basis consisting of eigenvectors of $A$. For the rest of this paper, we assume the following.
 
\begin{Assumption}\label{assumpt:nice}
The following are given:
\begin{enumerate}[(a)]
\item $A_G$ is a self-adjoint graph shift operator on $G$.
\item\label{assumpt:A} $A$ is a compact, self-adjoint and injective operator on $\mathcal{H}$. 
\item $\Phi$ is an orthonormal basis of $\bbC^n$ consisting of the eigenvectors of $A_G$. $\Xi$ is an orthonormal basis of $\mathcal{H}$ consisting of the eigenvectors of $A$.
\end{enumerate}
\end{Assumption}
Assumption~\ref{assumpt:nice}\ref{assumpt:A} may be further generalized in some cases to allow the operator $A$ to be different for different vertices of $G$. We discuss this generalization in Section~\ref{sec:ada}.

For $\phi \in \Phi$ and $\xi \in \Xi$, we use $\lambda_{\phi}$ and $\lambda_{\xi}$ to denote their corresponding eigenvalues. Now we can make the following definition. 

\begin{Definition} \label{defn:fft}
For $f\in S(G,\mathcal{H})$, the \emph{frequency range} of $f$ is defined to be $\{(\lambda_{\phi},\lambda_{\xi}^{-1}) \in \mathbb{R}\times \mathbb{R} : \mathcal{F}_f(\phi\otimes \xi)\neq 0, \phi\in\Phi,\ \xi\in\Xi\}$ for $\Phi$ and $\Xi$ in Assumption~\ref{assumpt:nice}.
\end{Definition}

We use $\lambda_\xi^{-1}$ in Definition~\ref{defn:fft}, which is more convenient when we deal with the notion of bandlimitedness in Section~\ref{subsec:ban}. We also motivate using $\lambda_{\xi}^{-1}$ instead of $\lambda_{\xi}$ in Example~\ref{eg:itcc}\ref{it:smi} below.

\begin{Example}\ \label{eg:itcc} 
\begin{enumerate}[(a)]
\item In the case of $\mathcal{H}=\bbC$, every operator $A(a) = Aa$, where $A\in\bbC\backslash\{0\}$, is compact, self-adjoint and injective. Then the frequency range of $f$ is $\{(\lambda_\phi, 1/A) : \ip{f}{\phi}_\bbC\ne 0,\ \phi\in\Phi\}$, which is equivalent to the frequency range in traditional GSP since $A$ is a constant.

\item \label{it:smi} Suppose $\mathcal{H}$ is the subspace of $L^2([0,2\pi])$ consisting of $f$ such that $f(0)+f(2\pi)=0$ (cf. Example~\ref{eg:main}\ref{it:2}). It is a closed subspace of $L^2([0,2\pi])$ and hence it is also a Hilbert space. Consider the differential operator $D = -\iu \dfrac{\ud}{\ud x}$. The eigenvectors of $D$ given by $\Xi = \{\xi_m=\exp(\iu(m+1/2)x)/\sqrt{2\pi} : m \in \mathbb{Z}\}$ forms an orthonormal basis of $\mathcal{H}$ where the eigenvalue of $\xi_m$ is $m+1/2$. Note that $\Xi$ is a variant of the standard Fourier basis. However, $D$ is not well-defined on all of $\mathcal{H}$. To fit into our framework, we define 
\begin{align*}
A(f)(x) = \frac{\iu}{2}\left(\int_0^{x}f(t) \ud t - \int_x^{2\pi}f(t) \ud t\right),
\end{align*}
for all $f \in \mathcal{H}$. It is compact, self-adjoint and injective, with eigenvectors $\Xi$ and corresponding eigenvalues $\{(m+1/2)^{-1}: m\in\bbZ\}$ because the composition $A\circ D$ is the identity map on the domain of $D$. 
%<*tag:isp>
Since $\Xi$ is a basis, we can write 
\begin{align*}
 f(x) = \sum_{m\in \mathbb{Z}}\frac{a_m}{\sqrt{2\pi}} e^{\iu(m+1/2)x},\  a_m \in \mathbb{C} \text{ for all } m\in\bbZ.
\end{align*}	
In traditional Fourier series, if the coefficient $a_m$ of $\xi_m=\exp(\iu(m+1/2)x)/\sqrt{2\pi}$ is non-zero, we say $m+1/2$ belongs to the frequency range of $f$. However, $m+1/2$ is an eigenvalue of $D$ and not $A$. Instead, due to the fact that $A\circ D$ is the identity map, $(m+1/2)^{-1}$ is an eigenvalue of $A$ with the same eigenvector $\xi_m$. Therefore, since we want to use eigenvalues of the compact operator $A$ to describe the notion of frequency, we take the reciprocal of $(m+1/2)^{-1}$, i.e., $m+1/2$, to be in the frequency range. This makes our definition consistent with definitions from traditional Fourier series.
%</tag:isp>
This example thus motivates our use of $\lambda_{\xi}^{-1}$ in Definition~\ref{defn:fft}. 

For a specific illustration, suppose that $G$ is the undirected path graph with three vertices $u_1,u_2,u_3$ shown in \figref{fig:9}. Let $A_G$ be its Laplacian matrix, whose eigenvalues and eigenvectors are given by:
\begin{align*}
\lambda_{\phi_1} = 0,&\ \phi_1=\frac{1}{\sqrt{3}}[1,1,1]^T,\\
\lambda_{\phi_2} = 1,&\ \phi_2=\frac{1}{\sqrt{2}}[0,-1,1]^T,\\
\lambda_{\phi_3} = 3,&\ \phi_3=\frac{1}{\sqrt{6}}[-2,1,1]^T.
\end{align*}
Consider $f \in S(G,\mathcal{H})$ that assigns $f(u_2,x)=2\cos(x/2)$ to the central node, and $f(u_1,x)=\sqrt{2}\sin(\frac{1}{2}x-\frac{\pi}{4})$ and $f(u_3,x)=\sqrt{2}\sin(\frac{1}{2}x+\frac{\pi}{4})$ to the side nodes. It can be verified that $\mathcal{F}_f(\phi\otimes \xi)\neq 0$ only for $\phi\otimes \xi \in\{\phi_2\otimes \xi_{-1}, \phi_2\otimes \xi_{0}, \phi_3\otimes \xi_{-1}, \phi_3\otimes \xi_{0}\}$. For example, by Lemma~\ref{lem:fap}, we obtain 
\begin{align*}
&\mathcal{F}_f(\phi_2\otimes \xi_{-1}) 
= \ip{\calG_f(\phi_2)}{\xi_{-1}}_\mathcal{H}\\
&= \int_0^{2\pi}  \frac{1}{\sqrt{2}}(0-2\cos\frac{x}{2}+\sqrt{2}\sin(\frac{x}{2}+\frac{\pi}{4}))\cdot \ofrac{\sqrt{2\pi}}e^{\iu x/2} \ud x \\  
&= -\frac{\sqrt{\pi}}{2} + \iu\frac{\sqrt{\pi}}{2}. 
\end{align*}
Hence, the frequency range of $f$ is $\{(1,-1/2),(1,1/2),(3,-1/2),(3,1/2)\}$.

\begin{figure}[!htb] 
	\centering
	\includegraphics[width=0.5\columnwidth]{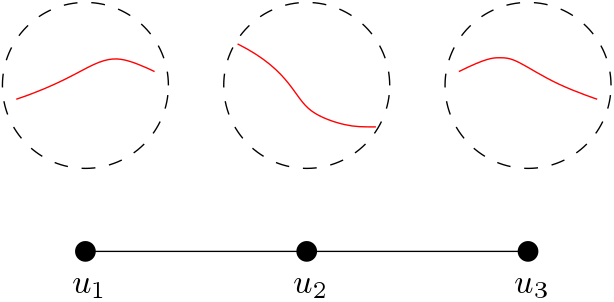}
	\caption{Illustration of Example~\ref{eg:itcc}\ref{it:smi}. Each red curve depicts the signal at each node of $G$, which is a function in $L^2([0,2\pi])$.} \label{fig:9}
\end{figure}
\end{enumerate}
\end{Example}

\section{Filtering} \label{sec:fil}
%<*tag:wcf>
In this section, we consider filtering of generalized graph signals. We focus on filters defined based on $\mathcal{F}$-transforms. Filters can have several different uses, to name a few: they can be used remove noise in signals, describe inherent relations between datasets and transform a signal into a different domain that is more convenient for analysis. Our generalized GSP filtering theory have parts similar to traditional GSP and signal processing over $\bbC$, while additional new features present due to the richer structure in a general $\mathcal{H}$. We discuss a few general families of filters that are related with each other and illustrate them with examples. We also point out why some of them are particularly useful.
%</tag:wcf>	

\begin{Definition}
A \emph{filter} is a bounded linear transformation $L:S(G,\mathcal{H}) \to S(G,\mathcal{H})$. 
\end{Definition}

Since $S(G,\mathcal{H})$ is a Hilbert space, any filter $L$ is continuous as it is bounded. From isomorphism, we equivalently regard any filter $L$ on $S(G,\mathcal{H})$ to be a filter on $\bbC^n\otimes\mathcal{H}$. 

%\begin{Example}\ \label{eg:lmb}
%\begin{enumerate}[(a)]
%\item\label{it:lmba} Let $m>0$ be a fixed integer. Suppose that at each node $v\in V=\{1,\ldots,n\}$, there is a graph $G_v$ with $m$ vertices. For different nodes $v$, the graphs $G_v$ might be different (cf. Example~\ref{eg:main}\ref{it:1}). The generalized signal at each node $v$ is defined as a graph signal on $G_v$. In this case, we can let $\mathcal{H}= \mathbb{C}^m$. Let $A_v$ be a filter of the graph signals on $G_v$, e.g., the Laplacian of $G_v$. Each $A_v$ can be viewed as an $m\times m$ matrix. 
%
%Suppose $M$ is a filter on $S(G,\bbC)$, which can be viewed as an $n\times n$ matrix. Let $M_i$ be the $n\times n$ matrix with $i$-th column the same as $M$ and the remaining columns being $0$. We are now able to construct a filter on $S(G,\mathcal{H}) \cong \mathbb{C}^{nm}$ as $\sum_{1\leq i\leq n}M_i\otimes A_i$ (here, $\otimes$ is the matrix Kronecker product). Clearly, if all the graphs $G_i$ are the same and filters $A_i=A$ are the same for $1\leq i\leq n$, the constructed filter is just $M\otimes A$.
%\item\label{it:shwh} Suppose $\mathcal{H} = L^2([a,b])$ (cf. Example~\ref{eg:main}\ref{it:2}). We have seen that $S(G,\mathcal{H})$ can be identified with $L^2(G\times [a,b])$, which is an infinite dimensional Hilbert space. Therefore, in the definition of a \emph{filter}, in addition to being linear, we require that it is also continuous, which is equivalent to being bounded. 
%\end{enumerate}
%\end{Example}

\subsection{Shift Invariant Filters}

For the two operators $A_G$ and $A$ in Assumption~\ref{assumpt:nice}, we can form their tensor product: $A_G\otimes A$ induced (linearly) by $A_G\otimes A(v \otimes h) = A_G(v)\otimes A(h)$. Note that $A_G\otimes A$ is an operator on the Hilbert space $S(G,\mathcal{H})\cong \bbC^n\otimes\mathcal{H}$. As both $A_G$ and $A$ are compact ($A_G$ is compact because all operators on the finite dimensional space $\bbC^n$ are) and self-adjoint, so is $A_G\otimes A$. The orthonormal basis $\Phi\otimes\Xi$ consists of the eigenvectors of $A_G\otimes A$.

We can define $A_G\otimes \Id$ and $\Id\otimes A$, and abbreviated as $A_G$ and $A$ if no confusion arises. If $\mathcal{H}$ is also finite dimensional, then $A_G\otimes A$ is the Kronecker product of matrices $A_G$ and $A$.

\begin{Definition}\label{def:shift_invariant}
A filter $L$ is called \emph{shift invariant} if both $A_G\circ L = L\circ A_G$ and $A\circ L = L\circ A$ hold for $A_G$ and $A$ in Assumption~\ref{assumpt:nice}. It is \emph{weakly shift invariant} if $(A_G\otimes A) \circ L =L \circ (A_G\otimes A)$. 
\end{Definition}

These concepts reduce to the traditional notion of shift invariance in GSP if $\mathcal{H}=\mathbb{C}$ given in \cite{Shu13} since $A$ is trivial in that case. The following is an example of a shift invariant filter that we will frequently refer to in the sequel.
\begin{Example}\label{ex:polynomial}
Let $P(x)=a_0 + a_1 x + \ldots + a_p x^p$ be a polynomial of degree $p < \infty$. Then, $P(A_G\otimes A)$ commutes with both $A_G$ and $A$ and is thus shift invariant. Polynomials on $A_G\otimes A$ give an important family of shift invariant filters, which can be used to approximate other filters due to the Stone-Weierstrass Theorem.
\end{Example}
Shift invariant filters also appear in a large array of applications such as data compression, customer behavior prediction \cite{San13} and machine learning models such as graph convolutional networks \cite{Def16, Kip16}, in which polynomials of the shift operators are used. 

In general, a weakly shift invariant filter is not necessarily shift invariant. For a simple example, let $
	A=B=
	\begin{bmatrix}
	1 & 0 \\
	0 & 2
	\end{bmatrix}.
$
Then 
\begin{align*}
A\otimes B = 
	\begin{bmatrix}
1 & 0 & 0 & 0 \\
0 & 2 & 0 & 0 \\
0 & 0 & 2 & 0 \\
0 & 0 & 0 & 4 \\
\end{bmatrix} \text{ and }
A\otimes \Id =
	\begin{bmatrix}
1 & 0 & 0 & 0 \\
0 & 1 & 0 & 0 \\
0 & 0 & 2 & 0 \\
0 & 0 & 0 & 2 \\
\end{bmatrix}.
\end{align*}
It is easy to see that as long as the top row, bottom row, left-most column and right-most column of $L$ are zero, then $L$ commutes with $A\otimes B$, i.e., $L$ is weakly shift invariant. However, as the second and third diagonal entries of $A\otimes \Id$ are distinct, some of these $L$s do not commute with $A\otimes \Id$. We next describe situations where these two notions are equivalent.

\begin{Proposition} \label{prop:ili}
Suppose $L$ is a filter on $S(G,\mathcal{H})$. 
\begin{enumerate}[(a)]
\item\label{it:si_wsi} If $L$ is shift invariant, then it is also weakly shift invariant. 
\item\label{it:basis} If $\Phi \otimes \Xi$ in Assumption~\ref{assumpt:nice} consists of eigenvectors of $L$, then $L$ is shift invariant.
\item \label{it:ili} The eigenspace of each eigenvalue $\lambda\ne 0$ of $A_G\otimes A$ is of finite dimension $m_{\lambda}$. If $m_{\lambda}=1$ for all $\lambda$, then a weakly shift invariant filter is shift invariant.  
\item Suppose $L$ is self-adjoint. Then $L$ is weakly shift invariant if and only if $L$ is shift invariant. 
\end{enumerate}
\end{Proposition}

\begin{IEEEproof}\
\begin{enumerate}[(a)]
\item Suppose $L$ is shift invariant. We verify that
\begin{equation*}
\begin{split}
(A_G\otimes A)\circ L & = (A_G\otimes \Id)\circ (\Id\otimes A)\circ L \\
& = (A_G\otimes \Id)\circ L \circ (\Id\otimes A) \\
& = L\circ (A_G\otimes \Id)\circ (\Id\otimes A) \\
& = L \circ (A_G\otimes A).
\end{split}
\end{equation*}

\item Since $\Phi \otimes \Xi$ is a basis, its vectors are also eigenvectors to both $A_G\otimes \Id$ and $\Id\otimes A$ and shift invariance of $L$ follows from Defintion~\ref{def:shift_invariant}.

\item As $A_G\otimes A$ is compact, from \cite[Chapter 21.2, Theorem 6]{Lax02}), for $\lambda \neq 0, m_{\lambda}$ is finite. If $m_{\lambda}=1$ for all $\lambda$, then all eigenvalues are non-zero and each eigenspace $V_{\lambda}$ is one dimensional. Suppose $L$ is weakly shift invariant. Then each eigenspace $V_{\lambda}$ of $A_G\otimes A$ is an invariant subspace of $L$, i.e., $L(V_{\lambda}) \subset V_{\lambda}$. As each $V_{\lambda}$ is one dimensional for each eigenvalue $\lambda$ and the basis vectors in $\Phi\otimes \Xi$ are the eigenvectors of $A_G\otimes A$, they are also the eigenvectors of $L$. Therefore, $L$ is shift invariant from \ref{it:basis}.

\item Suppose $L$ is weakly shift invariant. Then, from \cite[Chapter 28 Theorem 7]{Lax02}, the basis $\Phi\otimes \Xi$ contains the eigenvectors of $L$ and \ref{it:basis} shows that $L$ is shift invariant. The converse follows from \ref{it:si_wsi}.
\end{enumerate}
\end{IEEEproof}

If each eigenspace of $A$ is one dimensional, we hope the same holds of $A_G\otimes A$ so that by Proposition~\ref{prop:ili}\ref{it:ili}, all weakly shift invariant filters are shift invariant. We have the following result. 

\begin{Proposition} \label{prop:see}
Suppose that the graph $G$ has at least $3$ nodes. Furthermore, each edge weight of $G$ is chosen randomly according to a distribution absolutely continuous with respect to (w.r.t.) the Lebesgue measure. Consider $A_G$ and $A$ in Assumption~\ref{assumpt:nice}. If each eigenspace of $A$ is one dimensional, then the following holds with probability one:
\begin{enumerate}[(a)]
\item If $A_G$ is the adjacency matrix of $G$, then $A_G\otimes A$ is injective and each eigenspace of $A_G\otimes A$ has dimension $1$. 
\item If $A_G$ is the Laplacian matrix of $G$, then the eigenspace corresponding to the eigenvalue $0$ of $A_G\otimes A$ is isomorphic to $\mathcal{H}$. Moreover, for each eigenvalue $\lambda\neq 0$, the eigenspace has dimension $m_{\lambda}=1.$ 
\end{enumerate}
\end{Proposition}
\begin{IEEEproof}
The proof of this result is technical and can be found in Appendix~\ref{proof:prop:see}.
\end{IEEEproof}

If $A_G$ is the Laplacian matrix, the eigenspace $\mathcal{V}_0$ of $A_G\otimes A$ corresponding to the eigenvalue $0$ is infinite dimensional if $\mathcal{H}$ is infinite dimensional. If $L$ is weakly shift invariant, then the orthogonal complement $\mathcal{H}'$ of $\mathcal{V}_0$ is an invariant subspace of $L$. To overcome the infinite dimensionality of $\mathcal{V}_0$ so that Proposition~\ref{prop:ili} is still applicable, we may consider the restriction of $A_G\otimes A$ to $\mathcal{H}'$. In this case, from Proposition~\ref{prop:see}, each eigenspace of $A_G\otimes A$ on $\mathcal{H}'$ is one dimensional with probability one if the edge weights of $G$ are chosen randomly according to a probability density function.

We end this subsection by providing examples of filters when $\mathcal{H}$ is infinite dimensional.

\begin{Example}\label{ex:tensor_nonSI}
If $L$ can be decomposed as a tensor product $L=A_G'\otimes A'$, then $L$ is shift invariant if and only if $A_G'$ is shift invariant w.r.t.\ $A_G$ and $A'$ is shift invariant w.r.t.\ $A$. 

Similar to Example~\ref{eg:itcc}\ref{it:smi}, consider $\mathcal{H} = L^2([-1,1])$ and $A$ is the operator 
\begin{align*}
A(f)(x) = \int_{-1}^xf(t)dt - \int_x^1 f(t)dt.
\end{align*}
Let $A'$ be the operator on $\mathcal{H}$ defined as
\begin{align*}
A'(f)(x) = \left\{ \begin{array}{rcl}  f(x+1) & \mbox{for} & -1\leq x\leq 0 \\ 0 & \mbox{for} & 0<x\leq 1. \end{array} \right. 
\end{align*}
Although $A'$ looks like a shift, $L = A_G\otimes A'$ is not shift invariant w.r.t.\ $A_G\otimes A$. To see this, we verify that $A\circ A' \neq A'\circ A.$ Consider $f(x) = x$ on $[-1,1].$ It is easy to verify that $A\circ A'(f)(x) > 0$ for $0\leq x\leq 1$. However, $A'\circ A(f)(x) = 0$ for $0\leq x\leq 1$.
\end{Example}

As explained above, polynomials on $A_G\otimes A$ are shift invariant. However, due to infinite dimensionality, not every shift invariant filter is a polynomial filter as shown in the following example.
\begin{Example}
Consider $\mathcal{H}$ a infinite dimensional Hilbert space and $A$ is as in Assumption~\ref{assumpt:nice}. Suppose $a\in \mathbb{R}$ is a positive real number larger than all the eigenvalues of $A$. Then $A' = (1-a^{-1}A)^{-1}$ is a bounded linear transformation. It has a convergent power series expansion in $A$, hence $A'\circ A = A\circ A'$. If we let $L = A_G\otimes A'$, then $L$ is shift invariant. However, it is in general not a polynomial in $A_G\otimes A$. 

This does not happen if $\mathcal{H}$ is finite dimensional, as $1-a^{-1}A$ is invertible in the polynomial ring generated by $A$ (i.e., $A'$ is a polynomial in $A$), due to the existence of the minimal polynomial.  
\end{Example}

\subsection{Compact and Finite Rank Filters}\label{subsec:Compact_and_FR}

Similar to the definition of a compact operator on the Hilbert space $\mathcal{H}$, a filter $L$ on $S(G,\mathcal{H})$ is compact if its image of the closed unit ball has compact closure. A filter is of finite rank if its image is finite dimensional.

In Example~\ref{ex:polynomial}, if $\calH$ is infinite dimensional, the filter $P(A_G\otimes A)$ is shift invariant but not compact if the polynomial $P$ has a non-zero constant term $a_0$. If $P$ does not have any constant terms, then $P(A_G\otimes A)$ is a compact filter since $A_G\otimes A$ is compact. On the other hand, it is also easy to construct a compact filter that is non-shift invariant (see comment at beginning of Example~\ref{ex:tensor_nonSI}). 

\begin{Corollary}\ \label{cor:compact_finite}
\begin{enumerate}[(a)]
	\item\label{it:frcom} If $L$ is a finite rank filter, then it is a finite sum of compact filters.
	\item\label{it:comfr} If $L$ is a compact filter, then it is the limit (in operator norm) of finite rank filters. More precisely, suppose we give an ordering $\{w_1,\ldots, w_i,\ldots\}$ of $\Phi\otimes\Xi$ and define $L_i$ to be the projection of the image of $L$ to the finite dimensional subspace $S_i$ spanned by $\{w_1,\ldots, w_i\}.$ Then $L_i$ converges to $L$ in operator norm as $i\to\infty$. 
\end{enumerate}
\end{Corollary}
\begin{IEEEproof}
The claims \ref{it:frcom} and \ref{it:comfr} follow from \cite[Theorem~4.8.11]{DebMik:B05} and \cite[Theorem~4.4]{Con:B90}, respectively.
\end{IEEEproof}

Consequently, to understand a compact filter $L$, we may instead study the finite rank approximations $L_i$ as in Corollary~\ref{coro:ili}\ref{it:comfr}. Suppose $A_G\otimes A$ has no repeated eigenvalues and $L$ is shift invariant. Consider $S_i$ in the corollary and let $F_i=A_G\otimes A\restriction_{S_i}: S_i \mapsto S_i$ to be the restriction of $A_G\otimes A$ on $S_i$. Since $S_i$ is an invariant subspace of $L$, abusing notations, we use the same notation $L_i: S_i \mapsto S_i$ for the restriction of $L_i$ to $S_i$. Then, we have the standard observation as \cite{San13}: $L_i$ is a polynomial in $F_i$ with degree at most $\dim(S_i)-1$ \cite{HorJoh:B90}.

\subsection{Convolution Filters} \label{subsec:con}

Suppose we fix a $g\in S(G,\mathcal{H})$. For each $f\in S(G,\mathcal{H})$, the element $g*f$ defined by $\mathcal{F}_{g*f} = \mathcal{F}_g\mathcal{F}_f$ is an element of $S(G,\mathcal{H})$, i.e.,
\begin{align}\label{gconvf}
g*f = \sum_{\phi\otimes\xi} \mathcal{F}_g(\phi\otimes\xi)\mathcal{F}_f(\phi\otimes\xi)\cdot \phi\otimes\xi,
\end{align}
where $\sum_{\phi\otimes\xi} |\mathcal{F}_{g}(\phi\otimes \xi)|^2 <\infty$ since $g\in S(G,\mathcal{H})$. It is easy to verify that $g\,*$ satisfies 
\begin{align*}
g*(af+h) = ag*f+g*h
\end{align*}
for $a\in\bbC$, $f,g\in S(G,\mathcal{H})$. Moreover, $g\,*: S(G,\mathcal{H}) \to S(G,\mathcal{H})$ is a bounded map (bounded by $\sup_{\phi\otimes \xi}|\mathcal{F}_g(\phi\otimes \xi)| <\infty$). Therefore, $g\,*$ is a filter. We call it a \emph{convolution filter}. In the case $\mathcal{H}=\mathbb{C}$, we are in the situation of traditional GSP. The notion of convolution filter agrees with the one given in \cite{Shu13}. 

For each $f = \phi\otimes \xi$ with $\phi \in \Phi$ and $\xi \in \Xi$, it follows from definition that $g*f = \mathcal{F}_g(\phi\otimes \xi) \cdot \phi\otimes \xi$. Hence $\phi\otimes \xi$ is an eigenvector of $g\,*$ with eigenvalue $\mathcal{F}_g(\phi\otimes \xi)$. From Proposition~\ref{prop:ili}\ref{it:basis}, $g\,*$ is a shift invariant operator. Moreover, since $\sum_{\phi\otimes\xi} |\mathcal{F}_{g}(\phi\otimes \xi)|^2 <\infty$, $g\,*$ is also a \emph{Hilbert-Schmidt} operator \cite[Chapter 30.8]{Lax02}, which leads to the following corollary.

\begin{Corollary} \label{coro:ili}
The filter $L=g\,*$ is compact and is the limit (in operator norm) of finite rank filters. 
\end{Corollary}
\begin{IEEEproof}
As $L=g\,*$ is Hilbert-Schmidt, it is compact (cf.\ \cite[Chapter~30.8, Exercise~11(g)]{Lax02}). Therefore, from Corollary~\ref{cor:compact_finite}\ref{it:comfr}, it is the limit of finite rank filters.  
\end{IEEEproof}

If $\calH$ is infinite dimensional, a polynomial filter $P(A_G\otimes A)$ as in Example~\ref{ex:polynomial} is non-compact if it has a non-zero constant term $a_0$, and is thus not a convolution filter. This differs from traditional GSP where all shift invariant filters are convolution filters (when $A_G$ does not have repeated eigenvalues) since in traditional GSP, shift invariant filters are polynomials of $A_G$ \cite{San14} and $\calH=\bbC$ is finite dimensional.

\subsection{Bandlimited Signals and Band Pass Filters} \label{subsec:ban}

A signal $f\in S(G,\mathcal{H})$ is said to be \emph{bandlimited} if its frequency range (cf.\ Definition~\ref{defn:fft}) is a bounded subset of $\Real\times\Real$. For any set $K \subset \Real\times\Real$, we use $S_K(G,\mathcal{H})$ to denote the set of signals whose frequency range belongs to $K$. As a special example, if $G$ is a point and $\mathcal{H}= L^2([a,b])$, the notion of bandlimitedness agrees with its classical counterpart in the setting of Fourier series.

\begin{Lemma}\ \label{lem:fib}
\begin{enumerate}[(a)]
\item\label{it:bl-f} $f$ is bandlimited if and only if its frequency range is a finite set.
\item If $K$ is bounded, then $S_K(G,\mathcal{H})$ is a finite dimensional subspace of $S(G,\mathcal{H}).$
\end{enumerate}
\end{Lemma}
\begin{IEEEproof}
\begin{enumerate}[(a)]
	\item As $A$ is compact, the eigenvalues of $A$ is bounded and accumulate only at $0$ (cf.~\cite[Chapter 21.2 Theorem 6]{Lax02}). Therefore, the set $\{(\lambda_{\phi},\lambda_{\xi}^{-1})\}$ is a discrete subset of $\Real\times\Real$ and $f$ is bandlimited if and only if its frequency range is a finite set.
	\item The claim follows immediately from (a).
\end{enumerate}
\end{IEEEproof}

For each $f \in S(G,\mathcal{H})$ and $K \subset \Real\times\Real$, we define the \emph{band pass filter} as a projection 
\begin{align}
P_K(f) = \sum_{(\lambda_{\phi},\lambda_{\xi}^{-1})\in K}\mathcal{F}_f(\phi\otimes \xi) \cdot \phi\otimes \xi.\label{eq:PKf}
\end{align}
In the $\mathcal{F}$-transform domain, $P_K$ is nothing but multiplying with the characteristic function of $K$. We have the following observation regarding $P_K$. Note that a band-pass filter is \emph{not} convolutional if $K$ is not bounded.

\begin{Corollary} 
For any $K \subset \Real\times\Real$, $P_K$ is shift invariant. Furthermore, it is a convolution filter if and only if $K$ is bounded. 
\end{Corollary}
\begin{IEEEproof}
As each $\phi\otimes \xi \in \Phi\otimes\Xi$ is an eigenvector of $P_K$, it is shift invariant from Proposition~\ref{prop:ili}\ref{it:basis}. Moreover, it is a convolution filter if and only if the characteristic function of $K$ is square summable on the discrete set $\{(\lambda_{\phi},\lambda_{\xi}^{-1}): \phi\in \Phi, \xi\in \Xi\}$, i.e., $K$ being finite. By Lemma~\ref{lem:fib}\ref{it:bl-f}, $K$ is finite if and only if it is bounded. 
\end{IEEEproof}

Based on the discussions in this section, we encounter a few scenarios in which finite dimensional subspaces of $S(G,\mathcal{H})$ are useful. This motivates the next section on sampling, in which we use a collection of points on $V\times \Omega$ to describe these subspaces.

\subsection{Adaptive Polynomial Filters} \label{sec:ada}
We devote this last subsection to GSP with signals belonging to adaptive networks (Example~\ref{eg:main}\ref{it:1}) by extending Example~\ref{ex:polynomial} to allow different operators $A$ for different vertices in $G$. 

	Let $m>0$ be a fixed integer. Suppose that at each node $u\in V$, there is a graph $G_u$ with $m$ vertices. For different nodes $u$, the graphs $G_u$ might be different (cf. Example~\ref{eg:main}\ref{it:1}). The generalized signal at each node $u$ is defined as a graph signal on $G_u$. In this case, we can let $\mathcal{H}= \mathbb{C}^m$. Let $A_u$ be a filter of the graph signals on $G_u$, e.g., the Laplacian of $G_u$. Each $A_u$ can be viewed as an $m\times m$ matrix. A filter $M$ on $S(G,\bbC)$ can similarly be viewed as an $n\times n$ matrix for a fixed ordering of the vertices in $V$.
	
	We call a filter $F$ an \emph{adaptive polynomial filter} with respect to (w.r.t.) $M$ and $\{A_u\}_{u\in V}$ if there are polynomials $P_1$ and $P_2$ such that $F = \sum_{u\in V}P_1(M)_u\otimes P_2(A_u)$, where $P_1(M)_u$ has the same $u$-th column as $P_1(M)$ and $0$ elsewhere. Here, $\otimes$ is the matrix Kronecker product.
	
	%It can be verified that $F$ is shift invariant w.r.t.\ $M$, i.e., $F\circ M = M\circ F$. The degree of $F$ is defined as $\max(\deg(P_1),\deg(P_2))$.
	
	Consider now the special case where $M$ and $A_u$ are adjacency matrices or Laplacian matrices of the graphs $G$ and $G_u$, respectively. Furthermore, suppose both $P_1$ and $P_2$ are degree~$1$ polynomials. Then, for any signal $f \in S(G,\mathcal{H})$, the $i$-th component of $F(f)$ at a node $u$ takes contribution from the $j$-th component of the signal at node $v$ if $(u,v)$ is an edge in $G$ and $(i,j)$ is an edge in the graph $G_v$ at node $v$.

\begin{Example}\label{eg:ive}
\begin{figure}[!htb] 
\centering
\includegraphics[width=0.6\columnwidth]{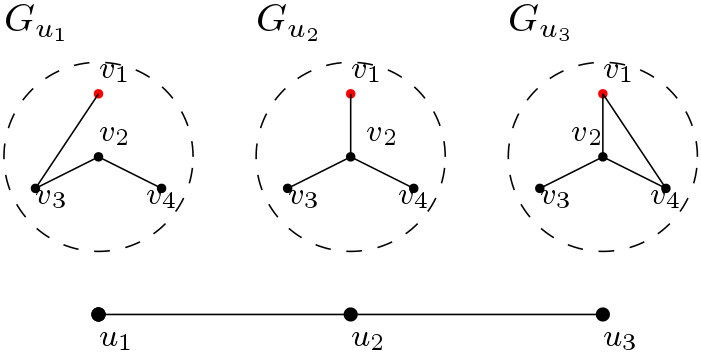}
\caption{Example to illustrate the action of the adaptive filter given in Example~\ref{eg:ive}. The $j$-th component of the graph signal at node $u$ contributes to the $i$-th component of the graph signal at node $v$, although $i$ and $j$ may not be connected by an edge in the graph at node $v$.} \label{fig:5}
\end{figure}

%<*tag:nwt>
Consider \figref{fig:5}. The graph $G$ is a path graph consisting of three nodes $u_1, u_2, u_3$. At each node $u_i$, the graph $G_{u_i}$ has $4$ nodes with different edge connections. Suppose $M$ is the adjacency matrix of $G$ and $A_{u_i}$ is the Laplacian matrix of $G_{u_i}$ for each $i=1,2,3$. Let $P_1(x) = a_1x+b_1$, $P_2(x) = a_2x+b_2$, and $F= \sum_{i=1}^3 P_1(M)_i\otimes P_2(A_{u_i})$ as above. Let $f \in S(G,\mathbb{C}^4)$. As an illustration, to evaluate $F(f)$ at $(u_2,v_1)$, we have 
\begin{align*}
F(f)(u_2,v_1) 
&= a_1\big( a_2(f(u_1,v_1)-f(u_1,v_3))+ b_2f(u_1,v_1)\big) \\ 
&+ b_1\big( a_2(f(u_2,v_1)-f(u_2,v_2))+b_2f(u_2,v_1)\big) \\ 
& + a_1\big( a_2(f(u_3,v_1)-f(u_3,v_2)+f(u_3,v_1)-f(u_3,v_4))+ b_2f(u_3,v_1)\big),
\end{align*}
i.e., $a_1$ weights the contribution from $G_{u_1}$ and $G_{u_3}$, $b_1$ weights the contribution from $G_{u_2}$, $a_2$ weights contribution of neighbors of $v_1$ in each $G_{u_i}$, and $b_2$ weights contribution of $v_1$ in each $G_{u_i}$. From this example, we see that the filter $F$ gives a weighted average of the signals in a neighborhood of each node in a neighborhood of graphs. 

From the above example, we see that an adaptive polynomial filter $F$ captures the hidden structures in $\mathbb{C}^4$ given by the graphs at each vertex. The modeling of such relationships is simplified by using the tools developed in Section~\ref{sec:gen} for representing generalized graph signals. Note that in practice, $G$ and each $G_{u_i}$ can have different physical meanings and scales (e.g., $G$ can be used to represent time while each $G_{u_i}$ represents the correlations between node observations at time instant $u_i$ in Example~\ref{eg:main}\ref{it:1}). It is inappropriate to then embed them in a big ambient graph and perform traditional GSP. This is an important reason for our proposed framework.
%</tag:nwt>
\end{Example}

\begin{figure}[!htb] 
	\centering
	\includegraphics[width=0.7\columnwidth]{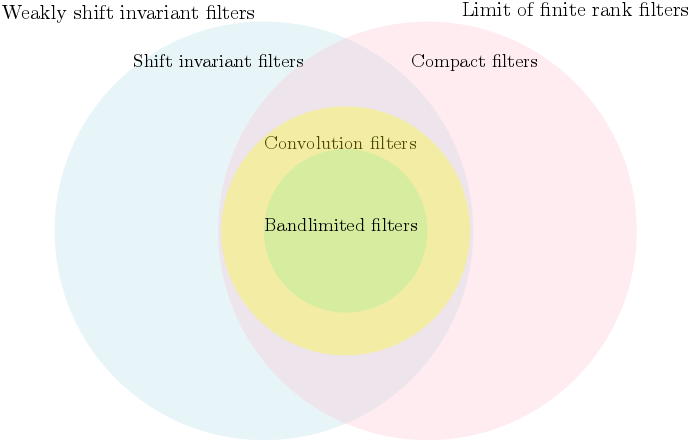}
	\caption{Summary of various filter families described in this section.} \label{fig:6}
\end{figure}

%<*tag:wst>
Finally, to conclude this section, we summarize the previous definitions of different filter families as a Venn diagram in \figref{fig:6}. Convolution filters and bandlimited filters are shift invariant filters that can be approximated by finite rank filters. Thus, they are particularly useful and they can be approximated by polynomials on $A_G\otimes A$, which can then be learned with an appropriate optimization procedure from observed signal samples.
%</tag:wst>

\section{Sampling} \label{sec:sam}
%<*tag:whb>
In the general setting of this paper, the Hilbert space $\mathcal{H}$ usually consists of certain functions on a domain $\Omega$. Sampling in this context has two stages: choose a subset of nodes $V'$ of $G$, and for each $v \in V'$, choose a finite subset from $\Omega$. The second stage can be both synchronous or asynchronous, depending on whether the sample set can be decomposed into a product $V' \times \Omega' $ for some finite subset $\Omega'\subset \Omega$. The need for asynchronous sampling is multi-folded, for example:

\begin{enumerate}[(a)]	
\item As we have explained in the introduction (cf. Example~\ref{eg:main}\ref{it:2}), it is not always easy to achieve synchronous sampling.

\item In the case of adaptive networks (Examples~\ref{eg:main}\ref{it:1} and \ref{eg:ive}), the Hilbert space $\mathcal{H}= \mathbb{C}^n$ is fixed, however, the coupled transform $A_v$ of $\mathcal{H}$ at each node $v$ of $G$ changes. Therefore, we may need to sample differently at different nodes of $G$.   
\end{enumerate}
Our generalized GSP framework allows us to develop a sampling theory that encompasses asynchronous sampling.
%</tag:whb>

In this section, we make additional assumptions regarding $\mathcal{H} = L^2(\Omega)$. We assume that $\Omega$ is a compact subset of a finite dimensional Euclidean space $\mathbb{R}^r$ for some $r\geq1$, and whose interior is non-empty and connected (for example, a finite closed interval), and $\mathcal{H}$ is equipped with the usual Lebesgue measure. As we pointed out earlier, each $f \in S(G,\mathcal{H}) \cong \mathbb{C}^n\otimes \mathcal{H}$ can be viewed as a function on $V\times \Omega$. A discussion of the case $\mathcal{H}=L^2(\Real)$ is deferred to Appendix~\ref{Appendix:L2R}.

Suppose $\mathcal{V}$ is a subspace of $S(G,\mathcal{H})$ with finite dimension $d_{\mathcal{V}}$. We want to choose a finite subset $W \subset V\times \Omega$ of size $d_{\mathcal{V}}$ such that each $f \in \mathcal{V}$ is uniquely determined by its values at $W$. We say that such a $W$ \emph{determines} $\mathcal{V}$. For each $v\in V$, let $k_v(W)$ be the number of points of $(\{v\}\times \Omega) \cap W$. 

To determine $\mathcal{V}$, $W$ cannot be formed in a completely random way. For example, in general, we cannot expect that choosing $d_{\mathcal{V}}$ points in $\{v\} \times \Omega$ for a single $v\in V$ does the job. However, we have a slightly weaker statement in Theorem~\ref{thm:sti}. Before stating the theorem, recall that a function $g: \Omega \to \mathbb{C}$ is \emph{analytic} if $g$ can be extended to a connected open neighborhood $U$ of $\Omega$ such that $g$ has a convergent Taylor series expansion in an open neighborhood of $x_0$ for any $x_0\in U$. A large family of common functions are analytic such as the polynomial functions, exponential functions and trigonometric functions.

\begin{Theorem}[Asynchronous sampling] \label{thm:sti}
Suppose $\mathcal{H}=L^2(\Omega)$ where $\Omega\subset\Real^r$ is compact, $\calV=V'\otimes\mathcal{H}'$ is a finite dimensional subspace of $S(G,\mathcal{H})$, where $V'\subset\bbC^n$, $\mathcal{H}'\subset\mathcal{H}$ is spanned by analytic functions, and $W$ determines $\mathcal{V}$. If $W'$ is formed by randomly choosing, according to a distribution absolutely continuous w.r.t.\ the Lebesgue measure, $k_v(W)$ points in $\{v\}\times \Omega$ for each $v\in V$, then $W'$ determines $\mathcal{V}$ with probability one.
\end{Theorem}
\begin{IEEEproof}
See Appendix~\ref{proof:prop:wcf}. 
\end{IEEEproof}

\begin{figure}[!htb] 
\centering
\includegraphics[width=0.7\columnwidth]{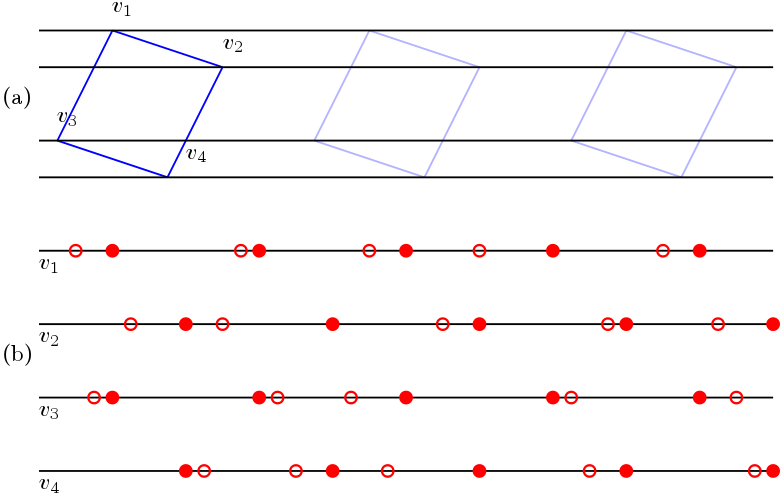}
\caption{Consider the graph $G=(V,E)$ with $4$ nodes and $4$ edges shown in (a) and suppose $\mathcal{H} = L^2([a,b])$. Suppose we want to construct $W$ that determines $\mathcal{V}=\text{span}(\Phi'\otimes \Xi')$ with $|\Phi'|=2$ and $|\Xi'|=10$. In (b), we show the domain $[a,b]$ for each vertex together with the sampling locations. By Example~\ref{eg:llb} and Corollary~\ref{prop:wcf}\ref{it:ksmall}, we can choose $W$ consisting of $5$ points for each $v\in V$ as shown by the red disks in (b). However, according to Theorem~\ref{thm:sti}, to determine $\mathcal{V}$, we can choose $W'$ consisting of red circles, which is a perturbation of $W$, and $W'$ almost surely determines $\calV$.} \label{fig:2}
\end{figure}

Intuitively, the theorem says that if one sampling scheme determines $\mathcal{V}$, then almost every other sampling scheme with the same number of sampled points at each vertex $v$ achieves the same effect. This observation makes particular use of the properties of $\Omega$. An illustration is shown in \figref{fig:2}.

Before proceeding further, we make the following definition.     

\begin{Definition} \label{defn:ckli}
Consider $k$ linearly independent vectors $\Psi=\{\psi_1,\ldots,\psi_k\}$ in $\mathbb{C}^n$ with $k\leq n$. For each subset $I \subset \{1,\ldots, n\}$, let $\psi_{i,I}\in \mathbb{C}^{|I|}$ be the vector formed by taking the components of $\psi_i$ indexed by $I$. Moreover, use $\Psi_I$ to denote $\{\psi_{1,I},\ldots, \psi_{k,I}\}.$ Define $\delta(\Psi)$ to be the largest integer $t$ such that there is a partition $\{1,\ldots, n\} = I_1 \cup \ldots \cup I_t$ into disjoint subsets, and each $\Psi_{I_j}, 1\leq j\leq t$ consists of linearly independent vectors.
\end{Definition}

Consider $\Phi'\subset \Phi$ of size $k$. Clearly, $\delta(\Phi')\leq \floor{n/k}$. In fact, it can be shown using a similar proof as that of Proposition~\ref{prop:see}, if the edge weights are chosen according to a distribution absolutely continuous w.r.t.\ the Lebesgue measure, then for any $k\leq n$ and $|\Phi'|=k$, $\delta(\Phi') =  \floor{n/k}$ with probability one. The choice of $I$ in Definition~\ref{defn:ckli} corresponds to a choice of vertices $V_I$ from $V$. Therefore, $\delta(\Phi')$ is the maximum number of disjoint subsets of vertices one can form so that $\Phi'_I$ has full column rank for each of the subsets $I$ (by definition, we have $|I| \geq |\Phi'|$ for each $I$), i.e, if the signals $\{f(v,x): v\in V_I\}$ at the vertices $V_I$ for $x\in\Omega$ are known, then $f(v,x)$ is uniquely determined for all $v\in V$.

\begin{Example} \label{eg:llb}
Let $A_G$ be the Laplacian matrix of an unweighted, undirected cycle graph with $4$ nodes. One finds an orthonormal basis $\Phi=\{(1,1,1,1),(-1,0,1,0),(0,-1,0,1),(-1,1,-1,1)\}.$ It is easy to verify the following from definition: for any subset $\Phi' \subset \Phi$	of size $2$, $\delta(\Phi')=2$. For example, if $\Phi' = \{(1,1,1,1),(-1,0,1,0)\}$, we can choose the partition $\{1,2,3,4\} = I_1\cup I_2$ with $I_1=\{1,2\}, I_2= \{3,4\}$. As a consequence, $\Phi_{I_1} = \{(1,1),(-1,0)\}$ and $\Phi_{I_2} = \{(1,1),(1,0)\}$, and both consist of linearly independent vectors.
\end{Example}

We consider the case where $\mathcal{V}$ is the span of $\Phi'\otimes \Xi'$, with $\Phi'$ and $\Xi'$ being finite subsets of $\Phi$ and $\Xi$ respectively. To perform sampling, we have two useful extreme cases: (a) we choose a small subset of $V' \in V$ and sample only at $\{v\} \times \Omega$ for $v\in V'$; (b) we sample at $\{v\}\times \Omega$ for each $v \in V$ with reduced amount of sample points. Case (a) corresponds to the situation where we make observations only at a small part of the graph; and case (b) corresponds to the situation where we make limited number of observations at each vertex. In the notation of Theorem~\ref{thm:sti}, we have the following result regarding the two sampling schemes (see also \figref{fig:2}). 

\begin{Proposition} \label{prop:wcf}
Assume the same conditions in Theorem~\ref{thm:sti}. Suppose $\Phi'$ and $\Xi'$ are finite subsets of $\Phi$ and $\Xi$ respectively and $\calV$ is the span of $\Phi'\otimes \Xi'$. 
\begin{enumerate}[(a)]
\item\label{it:necessity} Any $W$ that determines $\calV$ has $|W|\geq |\Phi'|\cdot|\Xi'|$.
\item \label{it:Vsmall} We can find $V'\subset V$ of size $|\Phi'|$ such that a set $W$ determines $\mathcal{V}$ with $k_v(W) = |\Xi'|$ for each $v\in V'$. For a fixed ordering of the vertices $V=\{v_1,\ldots,v_n\}$, a specific choice of $V'$ is given by $\{v_{i_1},\ldots,v_{i_{|\Phi'|}}\}$, which are the vertices corresponding to $|\Phi'|$ linearly independent rows of the matrix whose columns are formed by $\Phi'$.
\item \label{it:ksmall} We can find $W$ that determines $\mathcal{V}$ with $k_v(W)< |\Xi'|/\delta(\Phi')+1$ for each $v\in V$, and $\sum_{v\in V} k_v(W) = |\Xi'|$. %(see Fig~\ref{fig:10} for an illustration). 
\end{enumerate}
\end{Proposition}

%\begin{figure}[!htb] 
	%\centering
	%\includegraphics[width=0.75\columnwidth]{7}
	%\caption{Illustration of Proposition~\ref{prop:wcf}\ref{it:ksmall}. The sampling essentially uses an interlacing scheme.} \label{fig:10}
%\end{figure}

\begin{IEEEproof}
	See Appendix~\ref{proof:prop:wcf}. 
\end{IEEEproof}

Proposition~\ref{prop:wcf} together with Theorem~\ref{thm:sti} provides a simple sampling procedure: Proposition~\ref{prop:wcf}\ref{it:Vsmall} tells us how to choose a sample set of graph vertices $V'$ and $\{k_v(W): v\in V'\}$ while Proposition~\ref{prop:wcf}\ref{it:ksmall} allows us to choose $k_v(W)=\floor{|\Xi'|/\delta(\Phi')}+1$ for all $v\in V$ (in particular, if the edge weights of $G$ are randomly distributed according to a probability density function, we can choose $k_v(W)=\floor{|\Xi'|\cdot|\Phi'|/n}+1$). Then Theorem~\ref{thm:sti} allows us to randomly sample $k_v(W)$ points from $\Omega$ for each $v\in V'$ or $V$, using a probability density function. Note that Proposition~\ref{prop:wcf}\ref{it:ksmall} says that by utilizing the bandlimitedness of the underlying graph, we can use a sampling rate lower than the Nyquist rate to recover each graph vertex's signal from its samples. Consider Example~\ref{eg:itcc}\ref{it:smi}: if each $f(v,\cdot)$, $v\in V$, is bandlimited to a frequency band $[-B, B]$ in the classical Fourier series sense, then it is spanned by the set $\Xi' = \{\exp(\iu(m+1/2)x)/\sqrt{2\pi} : m+1/2 \in [-B,B]\}$ of eigenvectors. From the Shannon-Nyquist Theorem \cite{Sha49}, to recover $f(v,\cdot)$ for each $v$ individually, one needs at least $2B$ samples in $[0,2\pi]$. However, if we further know that $\{f(v,\cdot): v\in V\}$ is bandlimited in the graph dimension and spanned by $\Phi'$ with $\delta(\Phi')>1$, then a reduced number of samples ($\approx 2B/\delta(\Phi')$ for each vertex) is sufficient to recover all the vertex signals.

We remark that for applications like learning the polynomial form of a finite rank filter (cf.\ end of Section~\ref{subsec:Compact_and_FR}), choosing different sample sets corresponds to a base change. Therefore, it does not affect the coefficients of the polynomial. 

\section{Applications and Numerical Results} \label{sec:sim}

In this section, we discuss a few applications and present numerical results. A thorough discussion of each individual problem can be lengthy and thus beyond the scope of this paper.  We shall make a few simplifying assumptions, focus on how to apply the framework of the paper and demonstrate why the generalized GSP framework can be useful. In all the applications, we take $A_G$ to be the adjacency matrix and $\ip{\cdot}{\cdot}_{\bbC^n}$ to be the standard dot product in $\bbC^n$.

\subsection{Asynchronous Sampling}\label{subsec:asynchronous}

%<*tag:as>
Consider the asynchronous sampling given by the red circles in \figref{fig:2}. In that example, the generalized graph signal $f\in\text{span}(\Phi'\otimes \Xi')$ with $|\Phi'|=2$ and $|\Xi'|=10$. We see that although there are 20 samples in total, it is impossible to recover either $f(v,\cdot)$ or $f(\cdot,x)$ for each $v\in V$ and $x\in\calH$ individually. This makes it impossible to apply the time-vertex GSP framework as it requires uniform sampling. However, our asynchronous sampling results in Section~\ref{sec:sam} show that there are enough samples to recover the signal $f$. To do that, we require the generalized GSP framework introduced in this paper.

To illustrate this, we consider signal recovery from samples picked ``randomly''. More specifically, we choose two images (with the same size) of distinct digits. The graph $G$ is thus the $2D$-grid in which each vertex corresponds to a pixel. We take $\mathcal{H} = L^2([-1,1])$ with Chebyshev polynomials of the first kind $\{P_j\}_{j\geq 0}$ as the basis. A signal $f \in S(G,\mathcal{H})$ is chosen such that:
\begin{enumerate}[(a)]
	\item The graph signals $f(\cdot,1)$ and $f(\cdot,-1)$ correspond to the two chosen digit images.
	\item For each $x\in[-1,1]$, $f(\cdot,x)$ is graph bandlimited to the first $k=300$ eigenvalues of the graph Laplacian matrix. Furthermore, for each node $v$, the continuous signal $f(v,\cdot)$ is in the span of the first $B=8$ Chebyshev polynomials.
\end{enumerate}
Essentially, the signal $f$ depicts a smooth change from the first image to the second. Furthermore, we add white Gaussian noise with SNR$=10 B$ to obtain $\tilde{f}$.

\begin{figure}
	\centering
	\includegraphics[width=0.55\linewidth]{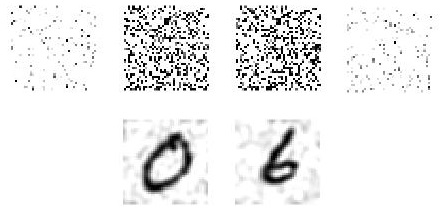}
	\caption{The top $4$ images are obtained from superimposed signals at the sample nodes and sample positions from $[-1,-0.5],[-0.5,0],[0,0.5],[0.5,1]$ respectively. The bottom two images are the recovered digits: $0$ and $6$.}\label{fig:10}
\end{figure}

According to Theorem~\ref{thm:sti} and Proposition~\ref{prop:wcf}, we can expect a recovery by sampling $2k$ nodes and $B/2$ random positions on $[-1,1]$ for each node. We chose the random positions following a normal distribution with mean $0$ and variance $0.5$. Thus, we expect samples to concentrate near $0$ and become sparse towards the end points $-1$ and $1$. Such a random procedure means that with probability one, for each $x\in [-1,1]$, at most one sampled pixel of $\tilde{f}(\cdot, x)$ is observed. We divide $[-1,1]$ into $4$ equal sub-intervals of size $0.5$ each, and superimpose the sampled signals for each interval. They are shown as the first $4$ images of \figref{fig:10}. Though the digits are barely observable from these samples, we indeed see that the two middle images carry more samples.

A basis of the space from which $f$ is chosen is $\{\phi_{i}\otimes P_j\}_{i\leq k, 0\leq j< B}$, where $\phi_i$ is an eigenvector of $A_G$ corresponding to the $i$-th smallest eigenvalue. Denote the samples by $W= (v_m, x_n)_{1\leq m\leq 2k, 1\leq B/2}$ and $\tilde{f}(W) = (\tilde{f}(v_m,x_n))_{1\leq m\leq 2k, 1\leq B/2}$. Then, each $\tilde{f}(v_m,x_n)$ is a noisy version of 
\begin{align*}
\sum_{i\leq k,0\leq j<B} y_{i,j}\phi_i(v_m)P_j(x_n)
\end{align*} 
for some $y = (y_{i,j})_{i\leq k, 0\leq j<B}$. Let $M$ be the corresponding transformation matrix with entries $\phi_i(v_m)P_j(x_n)$. We recover $y$ by solving the optimization:
\begin{align*} 
 \argmin_{y = (y_{i,j})_{i\leq k, 0\leq j<B}}\norm{My - \tilde{f}(W)}_2.
\end{align*}

The pixel values of the recovered images are obtained respectively as 
\begin{align*}
\sum_{i\leq k,0\leq j<B}y_{i,j}P_j(-1)\phi_i \text{ and } \sum_{i\leq k,0\leq j<B}y_{i,j}P_j(1)\phi_i.
\end{align*}
The recovered images are shown in \figref{fig:10} and we can see the digits $0$ and $6$ clearly.
%</tag:as>

\subsection{Network Information Propagation and Spectral Analysis}\label{subsec:inf_prop}

In this example, we illustrate the flexibility of generalized GSP over the time-vertex GSP proposed in \cite{Gra18}. The time-vertex GSP framework of \cite{Gra18} is briefly described in Example~\ref{eg:sts}\ref{it:swd}, and is equivalent to $\mathcal{H}=L^2(G')$, where $G'$ is a finite path graph, in our generalized GSP framework. This restricts signals at each vertex of $G$ to be a time series over a discrete set of time indices. Furthermore, to use the joint time-vertex Fourier transform (denoted as TV-transform for convenience) in \cite{Gra18}, the time index set of every vertex needs to be the same. 

In this example, we study graph signals generated from information propagation over a network $G$ (cf.\ \cite{Shah2011, LuoTayLeng13, LuoTay:C13, Zhu2016, JiTayVar:J17, Tan18}). Various infections spreading models have been considered under the independent cascade framework depending on whether an infected node can recover and become re-infected subsequently. For the Susceptible-Infected (SI) model, any infected node has a positive probability to infect its neighbors, and remains infected indefinitely. On the other hand, in the Susceptible-Infected-Recovered (SIR) model, an infected node has a probability to recover, afterwhich it cannot be infected again. Finally, in the Susceptible-Infected-Recovered-Infected (SIRI) model, any infected node can recover and become re-infected again. The difference in these spreading models results in different dynamical behaviors. For the SI model, all the nodes become infected almost surely; while for the SIR and SIRI models, it can happen that all the nodes are recovered if the recovery rate is high and infection rate is low. 

\begin{figure*}[!htbp]
\begin{subfigure}[b]{\linewidth}
\centering
\includegraphics[width=0.75\linewidth]{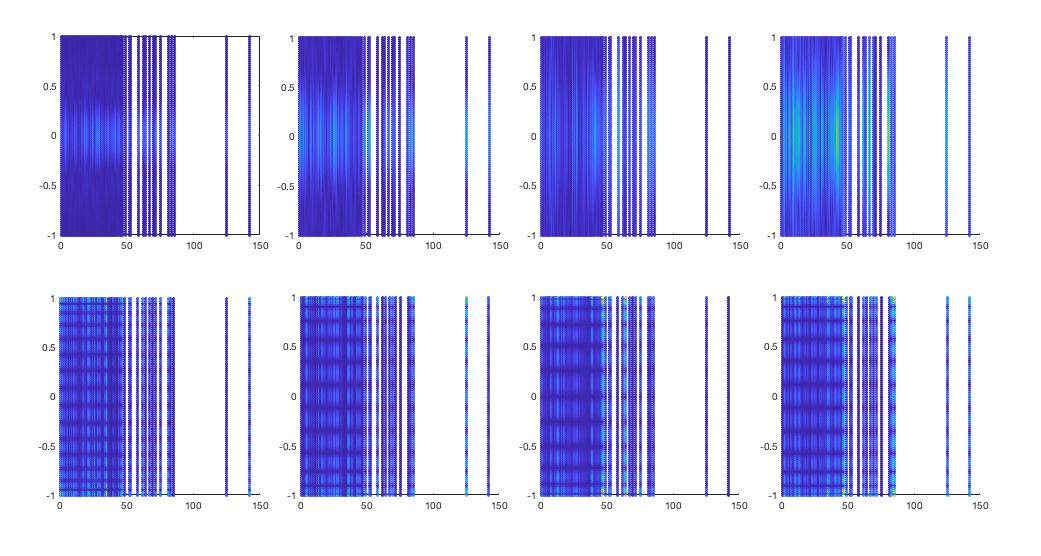}
\caption{Enron email network.}
\label{fig:3}
\end{subfigure}\\
\begin{subfigure}[b]{\linewidth}
\centering
\includegraphics[width=0.75\linewidth]{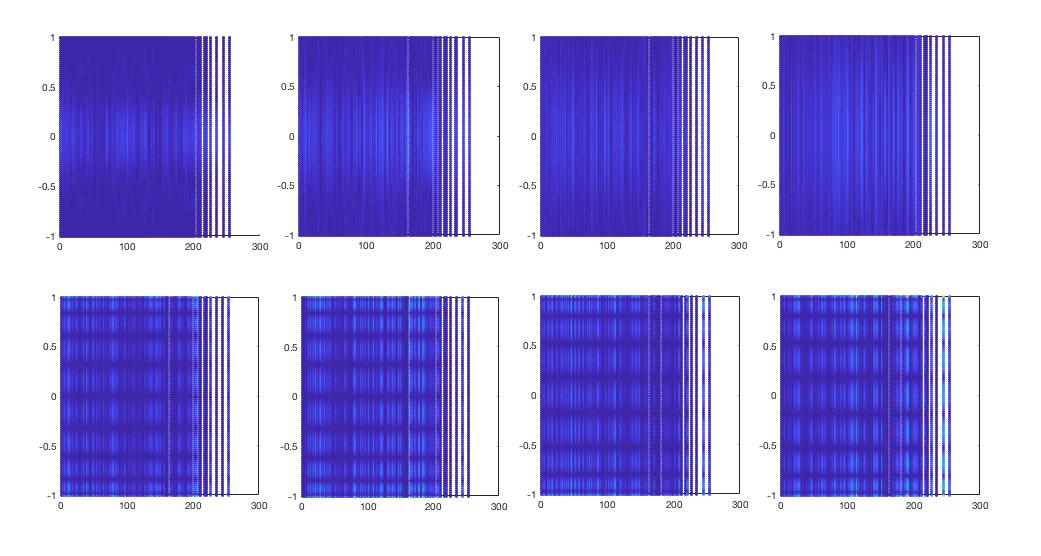}
\caption{Facebook network. }
\label{fig:4}
\end{subfigure}
\caption{The $\mathcal{F}$-transform (top row) and TV-transform (bottom row) spectrum plots for information cascade on the Enron email and Facebook networks. The horizontal axis shows the graph eigenvalues and the vertical axis shows the Fourier series frequencies. The plots correspond to (from left to right): SI model ($\lambda_I=1$), SIR model ($\lambda_I=1,\lambda_R=1/5$), SIRI model ($\lambda_I=1, \lambda_R=1/2$) and SIR model ($\lambda_I=1$, $\lambda_R=1$). A lighter color corresponds to a higher magnitude. }\label{fig:spectrum}
\end{figure*} 

%<*tag:fai>
Suppose that we use 0 to represent the uninfected status and 1 to represent the infected status at each node. Then the infection status of each node $v\in V$ follows a step function $f(v,\cdot)$, which is $L^2$ if we restrict to a finite observation period. The generalized GSP framework allows us to perform joint and partial $\mathcal{F}$-transforms on the generalized graph signal $f$ directly. On the other hand, to apply the time-vertex framework, one needs to perform uniform discrete sampling for each $f(v,\cdot)$, which may result loss of information since each $f(v,\cdot)$ is unbandlimited. Spectral analysis is a convenient mean to summarize signal features, and in the simulation below, we perform and compare spectral analysis of $f$ using both the $\mathcal{F}$-transform and TV-transform. 

For the simulation setup, we use all three information propagation models (SI, SIR, SIRI) over the Enron email network ($500$ nodes and average degree $12.6$) and a Facebook\footnote{https://snap.stanford.edu/data/egonets-Facebook.html} ($1034$ nodes and average degree $32.4$) network. The time-stamp of the occurrence of an infection or recovery event is generated using exponential distributions with means $\lambda_I$ or $\lambda_R$, respectively. Let $[0,T]$ be the time interval during which observations are made. For each node $v \in V$, we obtain a step function $f(v,\cdot)\in L^2([0,T])$. 

We compute the joint $\mathcal{F}$-transform and the TV-transform of $f$. For the $\mathcal{F}$-transform, we first compute the $\calH$-transform (partial $\mathcal{F}$-transform in the time direction), by noting that the Fourier transform of the standard rectangular function is the $\sinc$ function. For the TV-transform, we divide $[0,T]$ into uniform time slots and record the status of each node at the beginning of the slot as the graph signal for that time slot. Note that $f(v,\cdot)$ for each $v\in V$ is not a bandlimited signal in the time direction. Therefore, taking discrete samples of $f(v,\cdot)$ at a finite rate cannot recover the original graph signal. We plot the results in \figref{fig:3} and \figref{fig:4}, where the horizontal axis shows the graph eigenvalues and vertical axis shows the time-direction Fourier series frequencies. 

For a fixed $\lambda_I$, it is more difficult for the infection to spread across the network if $\lambda_R$ increases. Therefore, there is a gradual increase in difficulty in the diffusion process for the plots from left to right in \figref{fig:3} and \figref{fig:4}. If the diffusion process is fast, the initial spiky signal disappears fast in both the graph and time components. Therefore, we expect to see a smaller high intensity region. This agrees with what we see from \figref{fig:3} and \figref{fig:4}: we observe a clear spreading out of the higher energy part of the spectrum in the $\mathcal{F}$-transform as we go from the leftmost to the rightmost plot. This phenomenon is less discernible for the TV-transform. Moreover, for the TV-transform, a common spectral phenomenon is less obvious for different propagation types on the two networks. This is because the the $\mathcal{F}$-transform utilizes the full time series information at each graph vertex (i.e., the infection and recovery time stamps) whereas the TV-transform uses only the aggregated information in each discrete time slot.
%</tag:fai>

\subsection{Learning a Shift Invariant Filter} \label{sec:gts}

Consider $G$ with a time series over $[0,T]$ associated with each vertex. Suppose that the time interval $[0,T]$ can be divided into $q$ sub-intervals of equal size $T_0$. The graph time series has auto-regressive behavior from one sub-interval to the next via a shift invariant filter $F$. Therefore, under the generalized GSP framework, we consider $\mathcal{H} = L^2([0,T])$, and the time series for the $i$-th sub-interval is denoted as $f^{(i)}$. Our goal is to learn the filter $F$, where $f^{(i+1)} = F(f^{(i)})$, for $1\leq i < q$. 

We assume that the signals $\{f\tc{i}\}$ are bandlimited, i.e., there are numbers $B$ and $k$ such that each $f^{(i)}$ is bandlimited in the time direction by $B$, and the graph components are from the span of the first $k$ eigenvectors of the graph adjacency matrix $A_G$. We shall take the shift invariant filter $F=A_G\otimes P_2(L)$ where $P_2$ is a degree $2$ polynomial and $L$ is the translation by $T_0$ operator on the interval $[0,T]$ (with wrap around). We assume uniform sampling where the samples are chosen from a subset of vertices $V'$ according to Proposition~\ref{prop:wcf}\ref{it:Vsmall}. Let the samples in the $i$-th time sub-interval $[(i-1)T_0,iT_0]$ be denoted as a $|V'|\times B$ matrix $h\tc{i}$. We further add noise to each sample to obtain $\tilde{h}\tc{i}$. Our objective is to learn $P_2$ from $\{\tilde{h}\tc{i}\}$. 

We perform simulations on different graphs: random ER graphs ($1000$ nodes), the Enron Email graph\footnote{https://snap.stanford.edu/data/email-Enron.html} ($500$ nodes), and a synthetic company staff network ($80$ nodes, \cite{Pra18}). Let $T_0=2\pi$ and $q = 5$. For each experiment, we consider both $B=10$ and $B=20$, and set $k = 0.4|V|$. We randomly generate coefficients of the polynomial $P_2$ uniformly from $[0,1]$. The initial signal $f^{(1)}$ is generated as follows: we first generate real graph signals in the span of the first $k$ eigenvectors of $A_G$ with uniformly randomly chosen coefficients in $[0,1]$. Then, we use each component of these vectors as Fourier coefficients to generate a generalized signal at the corresponding vertex. We choose a subset of vertices $V'$ satisfying Proposition~\ref{prop:wcf}\ref{it:Vsmall} and sample $B$ time samples from the vertices in $V'$ in each sub-interval $[(i-1)T_0,iT_0]$, $1\leq i\leq q$. Independent Gaussian noise is then added to each sample $h\tc{i}$ to give a sequence of noisy (matrix) samples $\{\tilde{h}^{(i)}: 1\leq i \leq q\}$, from which we learn $P_2.$

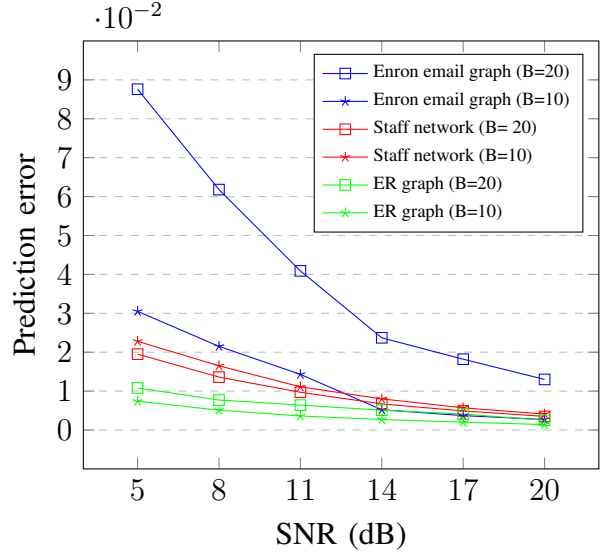
\begin{figure}[!htb]
	\centering
	\begin{tikzpicture}[scale=1]
	\begin{axis}[
	xlabel={SNR (dB)},
	ylabel={Prediction error},
	xmin=3, xmax=22,
	ymin=-0.01, ymax=0.1,
	scaled y ticks=base 10:2,
	xtick={5,8,11,14,17,20},
	ytick={0.00,0.01,0.02,0.03,0.04,0.05,0.06,0.07,0.08,0.09},
	legend pos=north east,
	legend style={font=\fontsize{7}{5}\selectfont},
	legend cell align={left},
	ymajorgrids=true,
	grid style=dashed,
	]
	
	\addplot[
	color=blue,
	mark=square,
	]
	coordinates {
		(5,0.0876)(8,0.0618)(11,0.0409)(14, 0.0237)(17,0.0182)(20,0.0130)
	};
	
	\addplot[
	color=blue,
	mark=star,
	]
	coordinates {
		(5,0.0305)(8,0.0215)(11,0.0143)(14, 0.0052)(17,0.0037)(20,0.0027)
	};
	
	\addplot[
	color=red,
	mark=square,
	]
	coordinates {
		(5,0.0195)(8,0.0136)(11,0.0097)(14, 0.0067)(17,0.0049)(20,0.0035)
	};
	
	\addplot[
	color=red,
	mark=star,
	]
	coordinates {
		(5,0.0228)(8,0.0165)(11,0.0111)(14, 0.0080 )(17,0.0057)(20,0.0041)
	};
	
	\addplot[
	color=green,
	mark=square,
	]
	coordinates {
		(5,0.0108)(8,0.0077)(11,0.0064)(14, 0.0051)(17,0.0041)(20,0.0025)
	};
	
	\addplot[
	color=green,
	mark=star,
	]
	coordinates {
		(5,0.0074)(8,0.0051)(11,0.0036)(14, 0.0027)(17,0.0020)(20,0.0014)
	};
	
	\legend{Enron email graph (B=20), Enron email graph (B=10),Staff network (B= 20), Staff network (B=10), ER graph (B=20), ER graph (B=10)}
	
	\end{axis}
	\end{tikzpicture}
	\caption{Plot of the prediction error against the SNR (in dB).} \label{fig:7}
\end{figure}

The recovery of $F$ preceeds in two steps (both involves an $L^2$ optimization): 
\begin{enumerate}[(a)]
	\item For the $t$-th sample time in $[(i-1)T_0,iT_0]$ for $1\leq i \leq q$, we recover the entire graph signal at that sample time from the samples of $V'$ as follows. Let $M$ be the matrix with $k$ columns corresponding to the sub-collection of eigenvectors of $A_G$ generating the signals. Let $M_{V'}$ be formed by taking rows corresponding to indices of $V'$. The graph signal $f\tc{i}(\cdot,t)$ is recovered by finding:
	\begin{align*}
	\tilde{f}\tc{i}(\cdot,t)= M\cdot\argmin_x \norm{M_{V'}x - \tilde{h}^{(i)}(\cdot,t)}_2,
	\end{align*} 
	where $h^{(i)}(\cdot,t)$ is the $t$-th column of $h\tc{i}$.
	\item Find $P_2$ by solving ($\norm{\cdot}_F$ below is the Frobenius norm):
	\begin{align*}
	\min_{P_2} \sum_{1\leq i\leq q}\norm{A_G\otimes P_2(L)(\tilde{f}\tc{i}) - \tilde{f}\tc{i+1}}_F^2.
	\end{align*}
\end{enumerate}

The performance is evaluated by computing the prediction error, i.e., 
\begin{align*}
\frac{\norm{(A_G\otimes \hat{P}_2(L))(h^{(m)})-h^{(m+1)}}_F}{\norm{h^{(m+1)}}_F},
\end{align*}
where $\hat{P}_2$ is the estimated $P_2$. In \figref{fig:7}, we plot the prediction error against the signal to noise ratio (SNR) in dB. We see that our procedure can learn $F$ well, and the performance improves with less noise.

\subsection{Adaptive Graph Signals}\label{subsec:adaptive}

In the third case study, we consider time series of graph signals with evolving graphs. In practice, a graph can evolve over time. Therefore even though a time series of graph signals belong to (the common) $\mathcal{H} = \mathbb{C}^n$, it can be inappropriate to disregard the underlying graphs that evolve over time. Application examples include sensor networks deployed in a dynamic environment like on the ocean surface, and social networks that evolve over time due to joining and leaving of users.

Consider Example~\ref{eg:ive}, where $G=(V,E)$ is a path graph representing the time line with $V=\{1,2,\ldots,n\}$ being the time indices, and each $G_t$ for $t\in V$ represents a graph with $m$ vertices at time $t$. With an initial graph $G_0$, we generate a sequence of graphs according to the evolution model proposed in \cite{Ito02}. For each $t=1,\ldots,n$, let $A_t$ be the adjacency matrix of $G_t$, which is assumed to be known. 

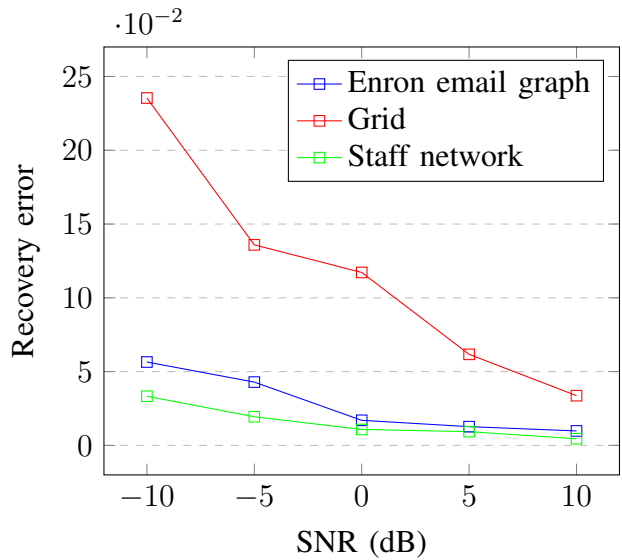
\begin{figure}[!htb]
\centering
\begin{tikzpicture}[scale=1]
\begin{axis}[
xlabel={SNR (dB)},
ylabel={Recovery error},
xmin=-12, xmax=12,
ymin=-0.02, ymax=0.27,
scaled y ticks=base 10:2,
xtick={-10,-5,0,5,10},
ytick={0.00,0.05,0.10,0.15,0.20,0.25},
legend pos=north east,
legend cell align={left},
ymajorgrids=true,
grid style=dashed,
]

\addplot[
color=blue,
mark=square,
]
coordinates {
(-10,0.0565)(-5,0.0429)(0,0.0170)(5, 0.0127)(10,0.0098)
};

\addplot[
color=red,
mark=square,
]
coordinates {
(-10,0.2354)(-5,0.1359)(0,0.1172)(5, 0.0617)(10,0.0337)
};

\addplot[
color=green,
mark=square,
]
coordinates {
(-10,0.0333)(-5,0.0194)(0,0.0108)(5, 0.0093)(10,0.0045)
};

\legend{Enron email graph, Grid, Staff network}

\end{axis}
\end{tikzpicture}
\caption{Plot of the recovery error against the SNR (in dB).} \label{fig:8}
\end{figure}

At each time or vertex $t$, we have a graph signal $f(t)\in\mathcal{H} = \bbC^m$, which we assume to be generated from a known base signal $g\in S(G,\mathcal{H})$ and a polynomial filter $F$, i.e., $f = F(g)$ in $S(G,\mathcal{H})$. The filter $F$ has the form $F = \sum_{1\leq t\leq n}P_1(A_G)_t\otimes P_2(A_t)$, where both $P_1$ and $P_2$ are degree~1 polynomials. In other words, there are parameters $a_0, a_1, b_0, b_1$ such that
\begin{align*}
f(t) = a_0\left(b_1A_tg(t)+b_0g(t)\right)+a_1\left(b_1A_{t-1}g(t-1)+b_0g(t-1)\right).
\end{align*}
In our experiments, we observe $\tilde{f}(t) = f(t) + N(t)$ for odd time indices, where $N(t)$ is an additive white Gaussian noise. Our objective is to infer $F$ (i.e, the parameters $a_0, a_1, b_0, b_1$) from the noisy samples $\{\tilde{f}(t): t\ \text{odd}\}$ by solving the optimization problem: 
\begin{align*}
\min \sum_{t\geq 3, \text{ odd}} \norm{F(g)(t) - \tilde{f}(t)}_2^2.
\end{align*}
The performance is evaluated by computing the recovery error of $f_1$ at even time slots: 
\begin{align*}
\sum_{t\geq 2, \text{ even}} \frac{\norm{\hat{F}(g)(t)-f(t)}_2}{\norm{f(t)}_2},
\end{align*}
where $\hat{F}$ is the estimated $F$. We perform simulations by choosing the initial graph $G_0$ to be the synthetic company staff network ($80$ nodes), grid ($400$ nodes), and the Enron Email graph ($500$ nodes).  We summarize the results in \figref{fig:8}. The optimization is non-convex, and may yield larger error with more noise. However, the procedure can learn $F$ well with less noise. 

\section{Conclusion} \label{sec:con}

In this paper, we have introduced the notion of graph signals in a separable Hilbert space, which we called generalized graph signals. We demonstrated how to define $\mathcal{F}$-transform as an analogy to the classical Fourier series and Fourier transform in traditional GSP. This leads to the notion of frequency. We developed theories on filtering and sampling, which find applications in reducing signals in infinite continuous domains to more manageable finite domains. We presented several scenarios in which the generalized GSP framework is more applicable than the traditional or time-vertex GSP frameworks.

The generalized GSP framework and its corresponding theory discussed in this paper is not only mathematically elegant but practical. We have only scratched the surface of utilizing this framework in different applications due to space constraint. In future work, it would be of interest to apply our framework to real datasets and to develop adaptations to different scenarios of interest. Another future direction involves developing a \emph{statistical} signal processing theory on top of the generalized GSP framework.

\appendices

\section{Proof of Proposition~\ref{prop:see}}\label{proof:prop:see}

\begin{enumerate}[(a)]
\item We first show that with probability one, $A_G$ is injective. We notice that $A_G$ is a symmetric matrix with $0$'s along the diagonal. Such matrices are parametrized by the upper triangular entries $\theta = (A_G(i,j))_{1\leq i < j\leq n}$. The determinant of $A_G$ is a (multi-variate) polynomial $M$ in $\theta$. As the adjacency matrix of the complete graph has non-zero determinant, the polynomial $M$ is not identically $0$. In view of the parametrization, the set $\mathcal{A}_0$ of adjacency matrices with zero determinant is a codimensional $1$ closed submanifold of the space of all adjacency matrices. Hence, the measure of $\mathcal{A}_0$, for any measure absolutely continuous w.r.t.\ the Lebesgue measure, is zero. This implies that $A_G\otimes A$ is injective with probability one since $A$ is assumed to be injective.

We next show that each eigenspace of $A_G\otimes A$ has dimension 1 with probability one. From Assumption~\ref{assumpt:nice} and the proposition hypothesis, the set of eigenvalues $\lambda(A)$ of $A$ is countable and each of its eigenspaces is one dimensional. Let $\mathcal{A}$ be the set of injective adjacency matrices $A_G$ such that $A_G\otimes A$ has repeated non-zero eigenvalues. It suffices to show the (product) Lebesgue measure, parametrized by the strict upper triangular entries, of $\mathcal{A}$ is $0$. We may further decompose $\mathcal{A} =\mathcal{A}_1\cup \mathcal{A}_2$ as follows:
\begin{enumerate}[(i)]
\item $A_G$ is in $\mathcal{A}_1$ if $A_G$ itself has repeated eigenvalues.
\item There are $\lambda_1,\lambda_2 \in \lambda(A)$, $\lambda_1\neq \lambda_2$ such that $\lambda_1A_G$ and $\lambda_2A_G$ has one common eigenvalue. The set of such $A_G$ is denoted by $\mathcal{A}_{2,\lambda_1,\lambda_2}.$ The set $\mathcal{A}_2 = \cup_{\lambda_1,\lambda_2} \mathcal{A}_{2,\lambda_1,\lambda_2}$ is a countable union.
\end{enumerate}

We show that both $\mathcal{A}_1$ and $\mathcal{A}_2$ have zero Lebesgue measure. In addition to the strict upper triangular entries $\theta$, we introduce one more parameter $\mu$ for an eigenvalue of $A_G$. In the case of $\mathcal{A}_1$, if $\nu$ is a repeated eigenvalue of $A_G$, then the parameters $(\theta,\nu)$ satisfy two conditions simultaneously:
\begin{enumerate}[(i)]
\item $P(\theta,\nu)=0,$ where $P=\det(\nu \Id - A_G)$ is the characteristic polynomial of $A_G$.
\item $P'(\theta,\nu)=0,$ where $P'$ is the partial derivative of $P$ w.r.t.\ $\nu$.
\end{enumerate}
As there are $A_G$ such that $P$ and $P'$ do not vanish simultaneously, the intersection of the two locus $P=0$ and $P'=0$ defines a codimension $\geq 1$ locus in the space of adjacency matrices (having $\mu$ adds one dimension and two independent relations remove two degrees of freedom). Therefore, $\mathcal{A}_1$ has Lebesuge measure $0$.

For $\mathcal{A}_2,$ we work with the countable union 
\begin{align*}
\bigcup_{\lambda_1\neq \lambda_2 \in \lambda(A)} \mathcal{A}_{2,\lambda_1,\lambda_2}.
\end{align*}
By countable subadditivity of measure, it suffices to show that each $\mathcal{A}_{2,\lambda_1,\lambda_2}$ has measure $0$. In this case, we use $\nu$ to parametrize the common eigenvalue of $\lambda_1A_G, \lambda_2A_G$. The parameters $(\theta, \nu)$ again satisfy $2$ conditions:
\begin{enumerate}[(i)]
\item $P(\theta,\nu/\lambda_1)=0,$ where $P$ is the characteristic polynomial.
\item $P(\theta,\nu/\lambda_2)=0.$ 
\end{enumerate}
We need to show that $P(\theta,\nu/\lambda_1)$ and $P(\theta,\nu/\lambda_2)$ are independent (the locus of one is not contained in the locus of the other). As above, we need to find $A_G$ such that $P(\theta,\nu/\lambda_1)$ and $P(\theta,\nu/\lambda_2)$ do not vanish simultaneously. We shall briefly indicate how such $A_G$ is chosen and omit the details. If $\lambda_1/\lambda_2 \neq -1$ and $n$ is even, $A_G$ can be chosen with $1$ at $A_G(i,n+1-i)$ and $0$ otherwise. If $\lambda_1/\lambda_2\neq -1$ and $n$ is odd, we let the top-left $3\times 3$ block being the adjacency matrix of the size $3$ complete graph and let the bottom-right $n-3 \times n-3$ block be chosen as the even case above. If $\lambda_1/\lambda_2=-1$, we choose three adjacency matrices (randomly choosing edge weights usually suffices) $A_{G_1}, A_{G_2}, A_{G_3}$ of size $3\times 3,4\times 4,5\times 5$ respectively, such that the absolute value of their eigenvalues are all distinct. Now for any $n\geq 3$, we can build up an $n\times n$ adjacency matrix by combining copies of $A_{G_1}, A_{G_2}$ and $A_{G_3}$ along the diagonal blocks. By our choice, $P(\theta,\nu/\lambda_1)=0$ and $P(\theta,\nu/\lambda_2)=0$ do not have common solutions. Therefore, by the same reasoning as above, the measure of $\mathcal{A}_{2,\lambda_1,\lambda_2}$ is $0$. This proves the claim.

\item By the similar argument as to (a) on the injectivity part (by considering the derivative of the characteristic polynomial at $0$), with probability one , the $0$-eigenspace $V_0$ of $A_G$ is one dimensional (consisting of vectors with the same entry in every row). Hence $V_0\otimes \mathcal{H} \cong \mathcal{H}$. The rest is a statement on the restriction of $A_G\otimes A$ to the orthogonal complement of $V_0$; and the proof is similar to (a).
\end{enumerate}

\section{Proofs of results in Section~\ref{sec:sam}}\label{proof:prop:wcf}

\subsection{Proof of Theorem~\ref{thm:sti}}
Let $W= \bigcup_{v\in V}\{(v,x_{v,t})\}_{1\leq t\leq k_v}$. Suppose we form $W'$ by randomly choosing $k_v=k_v(W)$ points in $\{v\}\times \Omega,$ denoted by $\{(v,y_{v,t})\}_{1\leq t\leq k_v}$. Let $k$ be the dimension of $\mathcal{V}$, which is not more than $\sum_vk_v.$ We may re-order and re-label $W=(w_l)_{1\leq l\leq k}=((v_l,x_l))_{1\leq l\leq k}$ and $W'=(w_l')_{1\leq l\leq k}=((v_l,y_l))_{1\leq l\leq k}$ such that $w_l$ and $w_l'$ belong to the same $\{v_l\}\times \Omega.$

Suppose $(g_i)_{1\leq i\leq k}$ forms a basis of $\mathcal{V}$, where for each $i=1,\ldots,k$, $g_i = \sum_j e_{i,j}\otimes f_{i,j}$, $e_{i,j} \in \mathbb{C}^n$, and $f_{i,j} \in \mathcal{H}$ are analytic for all $j$. Then each $f\in \mathcal{V}$ takes the form $\sum_{1\leq i\leq k}a_i g_i$.

The coefficient system $(a_i)_{1\leq i\leq k}$ is uniquely determined if and only if the $k\times k$ square matrix $M_{W'} = (g_i(w_l'))_{1\leq i,l\leq k}$ is invertible, i.e., it has non-zero determinant. For $W$, let $M_W=(g_i(w_l))_{1\leq i,l\leq k}$. 

Consider $w_l'=(v_l,y_l)$ and $g_i(w_l')=\sum_j e_{i,j}(v_l)f_{i,j}(y_l).$ The factor $e_{i,j}(v_l)$ is common for the $(i,l)$-th entries of both $M_W$ and $M_{W'}$. Therefore, the determinant $\det(M_{W'})$ is a polynomial in $\{f_{i,j}(y_l)\}$, and hence analytic in the variables $\{y_l\}$. It is known that, by analyticity, the following holds: the subset $Y\subset \Omega^k$ of $(y_l)_{1\leq l\leq k}$ making $\det(M_{W'}) = 0$ either has zero Lebesgue measure or $Y$ is the entire $\Omega^k$. However, the existence of $\{x_l\}_{1\leq l\leq k}$ (coming from $\{w_l\}_{1\leq l\leq k}$) shows that the latter case is not possible. As a consequence, $Y$ has zero measure for any measure absolutely continuous w.r.t.\ the Lebesgue measure.

\subsection{Proof of Proposition~\ref{prop:wcf}}
Claim \ref{it:necessity} follows directly from $f = \sum_{\Phi'\otimes\Xi'} \mathcal{F}_f(\phi\otimes\xi)\cdot\phi\otimes\xi$ for any $f\in\calV$.

We next show claims \ref{it:Vsmall} and \ref{it:ksmall}. Let $V=\{v_1,\ldots,v_n\}$, $1\leq t \leq \delta(\Phi')$, and $\{1,\ldots,n\}=I_1\cup\ldots\cup I_t$ a union of disjoint subsets such that each $\Phi'_{I_j}$ consists of linearly independent column vectors (cf.\ Definition~\ref{defn:ckli}). Therefore, $|I_j| \geq |\Phi'|$ for all $j=1,\ldots,t$. Since $\Phi'_{I_j}$ viewed as a $|I_j|\times |\Phi'|$ matrix contains at least $|\Phi'|$ linearly independent rows, we can choose row indices $i_l \in I_j$, $l=1,\ldots,|\Phi'|$, corresponding to linearly independent rows. Let $V_{I_j}= \{v_{i_1},\ldots,v_{i_{|\Phi'|}}\} \subset V$ be the corresponding graph vertices. Then, for any $x\in\spn(\Xi')$, $u\in V_{I_j}$ and $f\in \mathcal{V}$, we have
\begin{align*}
f(u,x) = \sum_{\phi\in\Phi'} \overline{\phi(u)} f(\phi,x),
\end{align*}
and $\{f(\phi,x):\phi\in\Phi'\}$ is uniquely determined by the values $\{f(u,x): u\in V_{I_j}\}$. Therefore, $f(v,x)$ is uniquely determined for all $v\in V$ since $f\in\calV = \spn(\Phi'\otimes\Xi')$. 

Since $|\Xi'|$ is finite, by a standard induction argument (identical to the proof that a $n\times k$ rank $k$ matrix has $k$ independent rows), we can  find $|\Xi'|$ points $\Omega' \subset \Omega$ such that for each $v\in V$ and $f(v,\cdot)\in \spn\Xi'$, $f(v,\cdot)$ is uniquely determined by its values at $\Omega'$. For any partition $\Omega' = \bigcup_{1\leq j\leq t}\Omega_j$ into $t$ disjoint subsets, we can construct $W$ as follows: if $i\in I_j,$ then $W\cap (\{v_i\}\times \Omega) = \{v_i\}\times \Omega_j$. Then, for each $x\in \Omega_j$, $v\in V$ and $f\in \mathcal{V}$, we have shown above that $f(v,x)$ is uniquely determined. As $\Omega' = \cup_{1\leq j\leq t}\Omega_j$, $f$ is uniquely determined.

The claims \ref{it:Vsmall} and \ref{it:ksmall} correspond to $t=1$ and $t=\delta(\Phi')$ respectively, where in case \ref{it:ksmall}, we choose $|\Omega_j|<|\Xi'|/t+1$ for each $j\leq t$.

\section{Square integrable functions of the line}\label{Appendix:L2R}

In this section, we present analogous discussion for the case $\mathcal{H} = L^2(\mathbb{R})$. In this case, the classical Fourier transform 
\begin{align}
f(v,x) \mapsto \widehat{f(v,\cdot)}(\omega) = \int_{-\infty}^{\infty} f(v,x)e^{-\iu \omega x}\ud x
\end{align}
for $f(v,\cdot) \in \mathcal{H}$ can no longer be viewed as a base change as the functions $e^{-\iu \omega x}$ are not square integrable (a unified approach requires the duality theory of locally compact abelian groups, which we would like to avoid). However, most parts of the theory can be developed in a parallel way. We will point out the differences in due course. 

Same as in Assumption~\ref{assumpt:nice}, $\Phi$ is an orthonormal basis consisting of the eigenvectors of the graph shift operator. For $f\in S(G,\mathcal{H})$, $\phi\in\Phi$ and $\omega\in\Real$, we may still define the joint $\mathcal{F}$-transform as 
\begin{align*}
\mathcal{F}_f(\phi, \omega) = \sum_{v\in V} \widehat{f(v,\cdot)}(\omega)\phi(v).
\end{align*}
and the partial $\mathcal{F}$-transforms as
\begin{align*}
\mathcal{H}_f(\omega)(v) &= \widehat{f(v,\cdot)}(\omega),\ \mathcal{G}_f(\phi)(x) = \langle f(\cdot,x),\phi\rangle_{\mathbb{C}^n}.
\end{align*}
so that
\begin{align*}
\mathcal{F}_f(\phi, \omega) = \ip{\mathcal{H}_f(\omega)(\cdot)}{\phi}_{\mathbb{C}^n} = \widehat{\mathcal{G}_f(\phi)}(\omega).
\end{align*}

We can similarly define the \emph{frequency range} of $f$ as  $\{(\lambda_\phi,\omega) : \mathcal{F}_f(\phi,\omega)\neq 0,\ \phi\in\Phi\}$, where $\lambda_\phi$ is the eigenvalue of $\phi$. It is \emph{bandlimited} if its frequency range is bounded. We can then define a band-pass filter by removing frequency components outside a designated frequency range.	To define convolution filters, we can make use of the convolution operator on $\mathcal{H}$. Suppose $g \in S(G,\mathcal{H})$ is such that $g(v) \in L^2(\mathbb{R})\cap L^1(\mathbb{R})$ for each $v\in V$ (e.g., if $g$ is compactly supported). For $f \in S(G,\mathcal{H})$, we define the convolution $g*f \in S(G,\mathcal{H})$ be such that 
\begin{align*}
\mathcal{F}_{g*f}(\phi,\omega) = \mathcal{F}_{g}(\phi,\omega)\mathcal{F}_{f}(\phi,\omega)
\end{align*}
for $\phi \in \Phi, \omega\in \mathbb{R}$. 

A discrete subset $W$ of $V\times \mathbb{R}$ is of \emph{sample rate} $r$ if for each connected interval $I\subset\Real$ of size $l$, the shifted interval $I_x = \{x+y : y\in I\}$ satisfy $|(V\times I_x)\cap W| = rl$ for all $x$ large enough. Suppose we have a \emph{closed} subspace $\mathcal{V}$ of $S(G,\mathcal{H})$, which is bandlimited. There is a bounded subset $K\subset\mathbb{C}\times \mathbb{R}$ such that the frequency range of each $f \in \mathcal{V}$ is contained in $K$. We say a countable discrete subset $W \subset V\times \mathbb{R}$ with positive sample rate determines $\mathcal{V}$ if each $f \in \mathcal{V}$ is uniquely determined by its values at $W$ as long as the values at $W$ are square summable. 

Let $l^2(W)$ be the space of square-summable sequences $(x_w)_{w\in W}$ indexed by $W$. We give $W$ the discrete topology and define the (evaluation) map 
\begin{align*}
e_{W}: \mathcal{V} \to l^2(W)
\end{align*} 
such that the component of the sequence $e_{W}(f)\in l^2(W)$ at the index $w=(v,x) \in W$ is $e_{W}(f)_w = f(v,x)$. Suppose $W$ determines $\calV$. Then, $e_W$ is injective. For each $w \in W$, let $\chi_w$ be the characteristic sequence indexed by $W$ that has value $1$ at $w$ and $0$ at $w'\in W \backslash \{w\}$. The \emph{standard basis} of $l^2(W)$ is given by $(\chi_w)_{w\in W}$. Let $f_w = e_W^{-1}(\chi_w)$. We say that $W$ has a \emph{rapidly vanishing} standard basis if for every pair $t,t' \in \mathbb{R}$ and $v\in V$, the sum $\sum_{w\in W}|f_w(v,t)-f_w(v,t')|^2<\infty$.

An example of a $W$ with a rapidly vanishing standard basis is $W$ consists of uniform samples at each vertex so that $f_w$, $w\in W$, are uniform translates of the $\sinc$ function. We have the following version of the asynchronous sampling theorem, which says by perturbing a finite subset of $W$ that determines $\mathcal{V}$ still determines $\mathcal{V}$. Changing finitely many sample points does not change the sample rate.

\begin{Theorem}[Asynchronous sampling for $\mathcal{H}=L^2(\Real)$]
Suppose $\calV\subset S(G,\mathcal{H})$ is a closed subspace that is bandlimited and $W$ determines $\mathcal{V}$ with a rapidly vanishing standard basis. Let $W_1$ be any finite subset of $W$. Let $W_2$ be $W$ such that each $w\in W_1$ is replaced by $r(w)$ such that both $w$ and $r(w)$ belong to $\{v\}\times \mathbb{R}$ for the same graph node $v\in V$, and $r(w)\ne r(w')$ for any $w\ne w'$. Then, $W_2$ determines $\mathcal{V}$.
\end{Theorem}
\begin{IEEEproof}
By the construction of $W_2$ from $W$, we have a bijection $W \to W_2$, $w \mapsto r(w)$ such that $w=r(w)$ for $w \notin W_1$.  Let $r(W_1) = \{r(w): w\in W_1\}$. By composing $e_{W_2}$ with $e_{W}^{-1}$, we obtain a linear map:
\begin{align*}
\psi= e_{W_2}\circ e_{W}^{-1}: l^2(W) \to l^2(W_2).
\end{align*}
To prove the theorem, it suffices to show that $\psi$ is invertible.

By assumption, $\{f_w = e_{W}^{-1}(\chi_w) : w\in W\}$ is rapidly vanishing. The image $\psi(\chi_w)=(f_w(w'))_{w'\in W_2}$ is a series that is $0$ except at $r(W_1)\cup \{r(w)\}$. It can be verified that: 
\begin{align*}
\norm{\psi(\chi_w)-\chi_{r(w)}}_2^2 = \sum_{x\in W_1} |f_w(r(x))-f_w(x)|^2.
\end{align*} 
We then have 
\begin{align*}
\sum_{w\in W}\norm{\psi(\chi_w)-\chi_{r(w)}}_2^2 = \sum_{w\in W}\sum_ {x\in W_1} |f_w(r(x))-f_w(x)|^2.
\end{align*} 
By our assumption, for each pair $x$ and $r(x)$, there is $v\in V$ such that $x=(v,t)$ and $r(x)=(v,t')$ with $t,t'\in \mathbb{R}$. Therefore, we have
\begin{align} 
\sum_{w\in W}\norm{\psi(\chi_w)-\chi_{r(w)}}_2^2 & = \sum_{w\in W}\sum_ {(v,t)\in W_1} |f_w(v,t')-f_w(v,t)|^2 \nonumber\\ 
& = \sum_{(v,t)\in W_1}\sum_{w\in W} |f_w(v,t')-f_w(v,t)|^2. \label{sumv}
\end{align}
Therefore, the right-hand side of \eqref{sumv} is finite as $W_1$ is finite. As $\{\chi_{r(w)} : r(w)\in W_2\}$ forms an orthonormal basis of $l^2(W_2)$, by a result due to Paley and Wiener \cite[Chapter~22.5, Theorem~7]{Lax02}, we know that $\{\psi(\chi_w): w\in W\}$ forms a basis of $l^2(W_2)$ and the proof is complete.
\end{IEEEproof}

\bibliographystyle{IEEEtran}
\bibliography{IEEEabrv,StringDefinitions,allref}

\end{document}